\documentclass[twocolumn]{openjournal}

\usepackage{tikz}
\usetikzlibrary{arrows.meta,positioning}
\usepackage{overpic}
\usepackage{graphicx}
\usepackage{rotating}
\usepackage{multirow}
\usepackage{natbib}
\usepackage{colorblind}
\usetikzlibrary{calc}
\usetikzlibrary{fit}
\usepackage{amsmath}
\usepackage{xcolor}

\defcitealias{saadting2025}{ST25}

\begin{document}

\title{No Gravitational Anomaly in Wide Binaries from Forward Modeling of 3D Orbits}

\author{Serat M. Saad}
\affiliation{Department of Astronomy, The Ohio State University, Columbus, OH 43210, USA}
\email{saad.104@osu.edu}

\author{Yuan-Sen Ting}
\affiliation{Department of Astronomy, The Ohio State University, Columbus, OH 43210, USA}
\affiliation{Center for Cosmology and AstroParticle Physics (CCAPP), The Ohio State University, Columbus, OH 43210, USA}
\affiliation{Max-Planck-Institut f{\"u}r Astronomie, K{\"o}nigstuhl 17, D-69117 Heidelberg, Germany}

\email{ting.74@osu.edu}

\begin{abstract}
Recent work by \cite{Chae2026} reported a gravitational anomaly in 36 wide-binary pairs, finding a gravity boost factor of $\gamma \equiv G_{\rm eff}/G_{\rm N} \approx 1.60_{-0.14}^{+0.17}$ at low accelerations, consistent with predictions from Modified Newtonian Dynamics (MOND). We reanalyze the same dataset using a hierarchical Bayesian model that infers a global $\gamma$ across all systems while fitting three-dimensional orbital elements. Our model yields $\gamma = 1.00^{+0.24}_{-0.19}$, consistent with Newtonian gravity ($\gamma = 1$). To identify the source of the discrepancy, we perform a test using an approach similar to \cite{Chae2026}, replacing the semi-major axis with a geometric de-projection of the observed projected separation. This test yields $\gamma = 1.56^{+0.21}_{-0.18}$, closely matching the result of \cite{Chae2026}. This shows that if the three-dimensional orbital separation is forward modelled using an independent semi-major axis parameter and then projected into the space of the observables, the relative velocities of wide binaries are consistent with Newtonian gravity.
\end{abstract}


\section{Introduction} \label{sec:intro}

Testing gravity in the low-acceleration regime is an ongoing challenge in astrophysics. On galactic scales, the observational predictions deviate from Newtonian predictions based on visible matter alone \citep{Zwicky1937, Kahn1959, Rubin1970, Faber1979, Ostriker1973, Sofue2001}. The standard $\Lambda$CDM cosmological model accounts for these deviations by invoking cold dark matter (CDM) \citep{Springel2005, Eisenstein2005}. The $\Lambda$CDM model has been successful at explaining large-scale structure and the cosmic microwave background \citep{Planck2020a, Planck2020b}. An alternative approach, Modified Newtonian Dynamics (MOND), was proposed by Mordehai Milgrom in 1983 \citep{Milgrom1983}. MOND suggests that gravity is modified below an acceleration scale $a_0 \approx 1.2 \times 10^{-10}~\rm{m\,s^{-2}}$ \citep{Milgrom1983}. MOND has been successful in reproducing galaxy rotation curves and scaling relations \citep{Sanders2002, Famaey2012, BanikZhao2022, Mcgaugh2016, Lelli2016}, and has been applied to the internal dynamics of elliptical galaxies \citep{Lelli2017} and dwarf galaxies in the Fornax Cluster \citep{Asencio2022}, but faces challenges on cluster and cosmological scales \citep{Sanders2003, Angus2008}, with recent constraints on Gpc scales from \citet{Russell2026}. Distinguishing between these two frameworks requires tests in environments where their predictions differ and the systematic uncertainties can be controlled. One such environment is the Solar System, where MOND has been tested using Cassini radio tracking of Saturn \citep{Hees2014, Park2026}, the associated constraint on the Solar System quadrupole \citep{Desmond2024}, and long-term orbital stability in the Kuiper Belt \citep{Vokrouhlicky2024}.

Wide-binary stars, separated by thousands of AU, provide another such environment, on a slightly larger but still sub-galactic scale, where gravity can be tested. At these separations, the internal gravitational acceleration of the binary can be compared to the MOND acceleration scale $a_0$. MOND framework therefore predicts deviation from Newtonian gravity in these systems. Testing MOND with wide binaries requires careful consideration of several theoretical ingredients. In MOND, the transition from Newtonian to modified gravity is governed by an interpolating function $\mu(x)$ where $x = a/a_0$, with $\mu \to 1$ for $x \gg 1$ (Newtonian regime) and $\mu \to x$ for $x \ll 1$ (deep-MOND regime). The exact form of $\mu$ is not uniquely specified by the theory, and different choices lead to different predicted gravity enhancements at intermediate accelerations \citep{Famaey2012}. Furthermore, wide binaries in the solar neighborhood are subject to the external field effect (EFE): the external gravitational acceleration from the Galaxy, $a_{\rm ext} \approx 1.8 \times 10^{-10}~\rm{m\,s^{-2}}$ at the solar position, is comparable to $a_0$ and much larger than the internal acceleration of the binary. In MOND, the EFE can suppress the internal anomaly, so the predicted gravity boost depends on $a_{\rm ext}/a_0$ as well as on the interpolating function and the specific MOND formulation \citep{Banik2018, Chae2022ext}. 

However, the gravity boost factor $\gamma \equiv G_{\rm eff}/G_{\rm N}$, defined as the ratio of the effective gravitational constant to the Newtonian value, can be measured independently of these theoretical choices. The inference of $\gamma$ is a kinematic measurement that assumes only Keplerian orbits; it does not assume any particular value of $a_0$, any specific interpolating function, or any MOND framework. Previous wide-binary studies using \textit{Gaia} data have taken this approach, reporting $\gamma = 1.43 \pm 0.06$ \citep{Chae2023}, $\gamma = 1.49 \pm 0.2$ \citep{Chae2024a}, $\gamma = 1.5 \pm 0.2$ \citep{Hernandez2024}, and most recently $\gamma = 1.6 \pm 0.2$ \citep{Chae2026}, which can be compared to AQUAL predictions of $\gamma \approx 1.4$ for wide binaries in the EFE-dominated regime of the solar neighborhood \citep{Banik2018, Chae2022ext}. In contrast, \cite{Banik2024} reported preference for Newtonian gravity ($\gamma = 1$). A recent study found that the claimed MOND signal in 2D wide-binary tests is largely attributable to sample quality control issues \citep{Cookson2026}. The conflicting conclusions highlight that the result may be sensitive to the statistical methodology or the data used for the analysis.

Recent work has introduced high-precision radial velocities (RVs) to enable three-dimensional analysis of wide-binary orbits. \cite{Saglia2025} obtained RVs for 32 wide binaries and found results broadly consistent with Newtonian gravity. \cite{Chae2025harps} reanalyzed the same sample and reported a tentative gravitational anomaly of $\gamma \equiv G_{\rm eff}/G_{\rm N} \approx 1.60_{-0.14}^{+0.17}$ for systems with low internal accelerations. Most recently, \cite{Chae2026} assembled a sample of 36 wide binaries with high-precision 3D velocities and reported $\gamma \approx 1.6 \pm 0.2$, claiming a $>3\sigma$ detection of deviation from Newtonian gravity. The methodology used by \cite{Chae2026} and \cite{Chae2025harps} consists of two steps: first, for each binary system, a Bayesian inference is performed to derive a posterior distribution of $\Gamma$, where $\Gamma = \log_{10} \sqrt{\gamma}$ along with nuisance orbital parameters; second, the per-system posteriors are statistically consolidated to derive a combined $\Gamma$. While this approach has merit, aspects of the model formulation, particularly the treatment of projected separation and the absence of an independent semi-major axis parameter, may influence the inferred $\gamma$.

In our previous work \citep[hereafter \citetalias{saadting2025}]{saadting2025}, we developed a hierarchical Bayesian framework for testing MOND with wide-binary kinematics and applied that to the wide binaries from the C3PO survey \citep{C3POI}. That model fit three-dimensional orbital elements for all systems while inferring a global MOND acceleration scale $a_0$. The framework included the MOND interpolating function and an analytic treatment of the EFE, and tested two interpolating functions ($b = 1$ and $b = 2$). The C3PO systems were found to reside in the MOND transition regime, and the analysis yielded tension with the canonical $a_0$ value at $\sim3\sigma$ ($b = 1$) and $\sim2\sigma$ ($b = 2$). Although, because the C3PO test was performed in the transition regime, the results were sensitive to the choice of interpolating function and the EFE treatment.

In this paper, we adapt the framework of \citetalias{saadting2025} to infer the gravity boost factor $\gamma$ for 36 wide binaries from \cite{Chae2026}. In Section~\ref{sec:data}, we describe the dataset from \cite{Chae2026}. In Section~\ref{sec:methods}, we present our hierarchical Bayesian model. In Section~\ref{sec:discussion}, we report the results of our two model variants, discuss the interpretation of these results, and place our findings in the context of previous wide-binary gravity tests.

\section{Data} \label{sec:data}

We use the sample of 36 wide-binary systems assembled by \cite{Chae2026}. This sample was constructed through selection criteria designed to identify gravitationally bound pairs with 3D kinematics: multi-epoch RV verification, speckle interferometry to exclude unresolved tertiary companions, Hipparcos-\textit{Gaia} proper motion consistency checks, and \textit{Gaia} RUWE $< 1.25$.

The RVs come from multiple instruments: the Las Cumbres Observatory (LCO) Network of Robotic Echelle Spectrographs (NRES), the Gemini-North MAROON-X spectrograph (MAROON-X), the High Accuracy Radial velocity Planet Searcher (HARPS) spectrograph on the European Southern Observatory (ESO) 3.6m telescope \citep{Saglia2025}, and the Apache Point Observatory Galactic Evolution Experiment (APOGEE) from Sloan Digital Sky Survey (SDSS) Data Release 17 (DR17) \citep{sdssdr17apogee}. The RV precisions vary across instruments, with HARPS providing the highest precision ($\sim$1--5~m\,s$^{-1}$) and LCO/NRES providing precisions of $\sim$50--300~m\,s$^{-1}$.


For each system, we use the following observational quantities: (i) the positions of both components from \textit{Gaia} DR3, from which we compute the projected separation and its uncertainty $\sigma_{r_\perp}$; (ii) the proper motions and their uncertainties from \textit{Gaia} DR3, from which we compute differential proper motions $\Delta\mu_\alpha$ and $\Delta\mu_\delta$; (iii) the RV difference between the two components and its uncertainty, as reported by \cite{Chae2026}; and (iv) stellar masses. For stellar masses, we use \textit{Gaia} FLAME estimates where available \citep{Gaiaflame}, supplemented by the mass estimates from \cite{Chae2023} for systems lacking FLAME values. Unlike our previous analysis in \citetalias{saadting2025}, which used differential RVs measured from the same spectrograph in a single epoch, the RVs in the \cite{Chae2026} sample are absolute RVs measured independently for each star, often from different instruments and epochs. The RV difference $\Delta v_r \equiv v_{r,B} - v_{r,A}$ is computed from these absolute measurements. All observational data are taken directly from \cite{Chae2026} and are available in the appendix of the paper. The angular separation is measured to \textit{Gaia} astrometric precision, so $\sigma_{r_\perp}$ is set mainly by the distance uncertainty and has a median fractional value of about $0.1\%$ across the sample. It is propagated through the likelihood, and Appendix~\ref{appendix:rproj_test} quantifies its effect on the inferred $\gamma$.


\section{Methods} \label{sec:methods}

Our model follows the hierarchical Bayesian framework developed in \citetalias{saadting2025}, with the key difference that we replace the MOND interpolating function and acceleration scale $a_0$ with a global gravity boost factor $\gamma$, defined such that $G_{\rm eff} = \gamma \cdot G_{\rm N}$. Newtonian gravity corresponds to $\gamma = 1$, while MOND predictions for wide binaries in the solar neighborhood give $\gamma \approx 1.4$--$1.6$ \citep{Banik2018, Chae2022ext, Chae2026}. The quantity $\Gamma = \log_{10}\sqrt{\gamma}$ used by \cite{Chae2026} is related by $\gamma = 10^{2\Gamma}$, with $\Gamma = 0$ corresponding to Newtonian.

We model each wide binary as a two-body Keplerian system with six orbital elements: semi-major axis $a$, eccentricity $e$, inclination $i$, longitude of ascending node $\Omega$, argument of periastron $\omega$, and mean anomaly $M$.

To keep the present analysis self-contained, we give the full mapping from the orbital elements to the observables. Given $M$, we solve Kepler's equation
\begin{equation}
M = E - e\sin E
\label{eq:kepler}
\end{equation}
for the eccentric anomaly $E$ by Newton--Raphson iteration. Following \citetalias{saadting2025}, we compute the separation and velocity directly in terms of $E$ rather than converting to the true anomaly $\nu$, which yields a simpler formulation. The instantaneous three-dimensional separation is
\begin{equation}
r_{\rm true} = a\,(1 - e\cos E),
\label{eq:rtrue}
\end{equation}
and the position of star B relative to star A in the orbital plane is
\begin{equation}
\mathbf{r}_{\rm orb} = \big(a(\cos E - e),\;\; a\sqrt{1-e^2}\,\sin E,\;\; 0\big).
\label{eq:pos}
\end{equation}

The velocity components in the orbital plane follow from differentiation with respect to time, using $\dot{E} = n/(1 - e\cos E)$ with mean motion $n = \sqrt{G_{\rm N} M_{\rm tot}/a^{3}}$:
\begin{equation}
\dot{\mathbf{r}}_{\rm orb} = \frac{n\,a}{1 - e\cos E}\,\big(-\sin E,\;\; \sqrt{1-e^2}\,\cos E,\;\; 0\big).
\label{eq:vel}
\end{equation}
In our parameterization the gravitational interaction is rescaled as $G_{\rm eff} = \gamma\,G_{\rm N}$, so the orbital velocity is scaled by $\sqrt{\gamma}$,
\begin{equation}
\mathbf{v}_{\rm orb} = \sqrt{\gamma}\;\dot{\mathbf{r}}_{\rm orb},
\label{eq:vmag}
\end{equation}
and the velocities of the two components in the orbital frame are
\begin{equation}
\mathbf{v}_{1,\rm orb} = \frac{M_2}{M_{\rm tot}}\,\mathbf{v}_{\rm orb},\qquad
\mathbf{v}_{2,\rm orb} = -\frac{M_1}{M_{\rm tot}}\,\mathbf{v}_{\rm orb}.
\label{eq:vcomp}
\end{equation}

We transform from the orbital plane to the ICRS frame with the rotation matrix
\begin{equation}
R = R_z(\Omega)\,R_x(i)\,R_z(\omega),
\label{eq:rot}
\end{equation}
giving $\mathbf{r}_{\rm ICRS} = R\,\mathbf{r}_{\rm orb}$ and $\mathbf{v}_{i,\rm ICRS} = R\,\mathbf{v}_{i,\rm orb}$. A common systemic velocity $\mathbf{v}_{\rm sys}$, which we marginalize over, is added to each component.

We then project onto the sky. At a position $(\alpha,\delta)$ the radial, east, and north unit vectors are
\begin{align}
\hat{\mathbf{r}} &= (\cos\delta\cos\alpha,\;\cos\delta\sin\alpha,\;\sin\delta),\nonumber\\
\hat{\mathbf{e}} &= (-\sin\alpha,\;\cos\alpha,\;0),\label{eq:triad}\\
\hat{\mathbf{n}} &= (-\sin\delta\cos\alpha,\;-\sin\delta\sin\alpha,\;\cos\delta).\nonumber
\end{align}
The projected separation is $r_\perp = \sqrt{r_E^2 + r_N^2}$, with $r_E = \mathbf{r}_{\rm ICRS}\cdot\hat{\mathbf{e}}_A$ and $r_N = \mathbf{r}_{\rm ICRS}\cdot\hat{\mathbf{n}}_A$. The differential radial velocity is
\begin{equation}
\Delta v_r = (\mathbf{v}_{2,\rm ICRS}+\mathbf{v}_{\rm sys})\cdot\hat{\mathbf{r}}_B - (\mathbf{v}_{1,\rm ICRS}+\mathbf{v}_{\rm sys})\cdot\hat{\mathbf{r}}_A.
\label{eq:dvr}
\end{equation}
For the proper motions, the tangential velocity of each component is $\mathbf{v}_{i,t} = (\mathbf{v}_{i,\rm ICRS}+\mathbf{v}_{\rm sys}) - [(\mathbf{v}_{i,\rm ICRS}+\mathbf{v}_{\rm sys})\cdot\hat{\mathbf{r}}_i]\,\hat{\mathbf{r}}_i$, and the differential proper motions are
\begin{equation}
\begin{aligned}
\Delta\mu_\alpha &= -\frac{1}{\kappa}\!\left(\frac{\mathbf{v}_{2,t}\cdot\hat{\mathbf{e}}_B}{d_B} - \frac{\mathbf{v}_{1,t}\cdot\hat{\mathbf{e}}_A}{d_A}\right),\\[2pt]
\Delta\mu_\delta &= \phantom{-}\frac{1}{\kappa}\!\left(\frac{\mathbf{v}_{2,t}\cdot\hat{\mathbf{n}}_B}{d_B} - \frac{\mathbf{v}_{1,t}\cdot\hat{\mathbf{n}}_A}{d_A}\right),
\end{aligned}
\label{eq:dpm}
\end{equation}
where $\kappa = 4.74047~\mathrm{km\,s^{-1}}\,(\mathrm{mas\,yr^{-1}\,kpc})^{-1}$ and $d_A,d_B$ are the component distances. Equations~\ref{eq:kepler}--\ref{eq:dpm} constitute the complete forward model, mapping the per-system parameters $\{a,e,i,\Omega,\omega,M,M_1,M_2\}$ and the global gravity boost factor $\gamma$ to the four predicted observables $\{r_\perp,\Delta v_r,\Delta\mu_\alpha,\Delta\mu_\delta\}$. The only change relative to \citetalias{saadting2025} is that the orbital velocity (Equation~\ref{eq:vmag}) carries the factor $\sqrt{\gamma}$ in place of the MOND velocity rescaling.

\subsection{Hierarchical Bayesian model} \label{sec:hiermodel}

\begin{figure*}
    \centering
    {\usetikzlibrary{fit,positioning}

\begin{tikzpicture}[
    scale=1.3,
    shorten >=1pt,->,draw=black!70,
    neuron/.style={circle,minimum size=40,inner sep=0},
    param/.style={neuron,fill=cyan}, 
    inter/.style={neuron,fill=black!40}, 
    gauss/.style={neuron,fill=red}, 
    obs/.style={neuron,fill=yellow}, 
    beta/.style={neuron,fill=yellow}, 
    box/.style={rectangle,draw=black,rounded corners,thick,inner sep=10pt}
]

\node[param] (a0) at (-0.75,5.5) {$\gamma$};
\node[gauss] (ua0) at (0.75,5.5) {$\mathcal{U}$};

\node[param] (a)   at (0,4) {$a$};
\node[param] (alpha)   at (-1.5,4) {$\alpha$};
\node[inter] (nu)  at (-3,1) {$E$};
\node[param] (M)   at (-3,4) {$M$};
\node[param] (i)   at (3,4) {$\cos i$};
\node[param] (omega)   at (4.5,4) {$\omega$};
\node[param] (Omega)   at (4.5,2.5) {$\Omega$};

\node[param] (M1M2) at (1.5,4)
  {\parbox{1cm}{\centering $M_1$ \\ $M_2$}};

\node[inter] (e) at (-1.5,1) {$e$};
\node[inter] (rtrue) at (-1.5,-0.5) {$r_{\rm true}$};

\node[inter] (R) at (3.75,1) {$R$};
\node[inter] (vproj) at (3.75,-0.5) {$\parbox{1cm}{\centering $v_{1,\perp}$ \\ $v_{2,\perp}$}$};

\node[inter] (v1v2) at (0.75,1) {$\parbox{1cm}{\centering $v_1$ \\ $v_2$}$};
\node[inter] (rv) at (0.75,-2) {$\Delta \rm RV$};
\node[inter] (pmra) at (2.25,-2) {$\Delta \mu_{\alpha}$};
\node[inter] (pmdec) at (3.75,-2) {$\Delta \mu_{\delta}$};
\node[inter] (rproj)   at (-1.5,-2) {$r_{\perp}$};

\node[gauss] (uM)   at (-3,2.5) {$\mathcal{U}$};
\node[gauss] (ue)   at (-1.5,2.5) {$\mathcal{U}$};
\node[gauss] (ua)   at (0,2.5) {$\mathcal{U}$};
\node[gauss] (gM1M2)   at (1.5,2.5) {$\mathcal{N}$};
\node[gauss] (uR)   at (3,2.5) {$\mathcal{U}_{3}$};
\node[gauss] (gv1v2) at (2.25,-3.5) {$\mathcal{N}_{4}$};

\node[obs] (robs) at (-0.75,-5) {$r_{\perp}$};
\node[obs] (rvobs) at (0.75,-5) {$\Delta \rm RV$};
\node[obs] (pmraobs) at (2.25,-5) {$\Delta \mu_{\alpha}$};
\node[obs] (pmdecobs) at (3.75,-5) {$\Delta \mu_{\delta}$};

\draw[->] (i) -- (uR);
\draw[->] (omega) -- (uR);
\draw[->] (Omega) -- (uR);
\draw[->] (uR) -- (R);
\draw[->] (R) -- (rproj);
\draw[->] (M) -- (uM);
\draw[->] (uM) -- (nu);

\draw[->] (a) -- (ua);
\draw[->] (ua) -- (rtrue);
\draw[->] (e) -- (rtrue);
\draw[->] (alpha) -- (ue);
\draw[->] (ue) -- (e);
\draw[->] (nu) -- (rtrue);

\draw[->] (M1M2) -- (gM1M2);
\draw[->] (gM1M2) -- (v1v2);
\draw[->] (rtrue) -- (rproj);

\draw[->] (a0) -- (ua0);
\draw[->] (ua0) -- (v1v2);
\draw[->] (rtrue) -- (v1v2);

\draw[->] (v1v2) -- (vproj);
\draw[->] (R) -- (vproj);
\draw[->] (vproj) -- (rv);
\draw[->] (vproj) -- (pmra);
\draw[->] (vproj) -- (pmdec);

\draw[->] (rproj) -- (gv1v2);
\draw[->] (gv1v2) -- (robs);
\draw[->] (rv) -- (gv1v2);
\draw[->] (gv1v2) -- (rvobs);
\draw[->] (pmra) -- (gv1v2);
\draw[->] (gv1v2) -- (pmraobs);
\draw[->] (pmdec) -- (gv1v2);
\draw[->] (gv1v2) -- (pmdecobs);

\node[box, fit=(a)(e)(M)(nu)
             (M1M2)(rtrue)(i)(omega)(Omega)
             (v1v2)(rv)(pmra)(pmdec)
             (gv1v2)(robs)(rvobs)(pmraobs)(pmdecobs)] (bigbox) {};

\node[anchor=south west,font=\small] at ([xshift=10pt,yshift=10pt]bigbox.south west) {For each pair};

\node[anchor=east] at ([xshift=-0.5cm]bigbox.west) {
    \begin{tikzpicture}[every node/.style={anchor=west}, x=1cm, y=0.6cm]
        \node[param,label=right:{Parameter}] at (3,6) {};
        \node[inter,label=right:{Intermdeiate Variables}] at (3,3) {};
        \node[gauss,label=right:{Distribution}] at (3,0) {};
        \node[obs,label=right:{Observed Variables}] at (3,-3) {};
    \end{tikzpicture}
};

\end{tikzpicture}}
    \caption{Graphical representation of the hierarchical Bayesian model for inferring the gravity boost factor $\gamma$. The global parameter $\gamma$ (blue) is shared across all binary systems. The plate notation indicates that the enclosed structure is replicated for each of the $N = 36$ systems. For each system, we infer six orbital elements ($a$, $\alpha$, $M$, $i$, $\omega$, $\Omega$) and masses ($M_1$, $M_2$). These determine intermediate quantities (gray): eccentric anomaly $E$, true separation $r_{\rm true}$, rotation matrix $R$, and component velocities $v_1$, $v_2$. The model predicts four observables (yellow): projected separation $r_\perp$, RV difference $\Delta v_r$, and differential proper motions $\Delta\mu_\alpha$, $\Delta\mu_\delta$. Red nodes denote probability distributions.}
    \label{fig:model_diagram}
\end{figure*}

Rather than fitting $\gamma$ independently for each system and then combining the results as in \cite{Chae2026}, we fit a global $\gamma$ shared across all 36 systems. The graphical model is shown in Figure~\ref{fig:model_diagram}.

The joint posterior is,
\begin{equation}
p(\gamma, \{{\theta}_j\} | \{{d}_j\}) \propto p(\gamma) \prod_{j=1}^{N} p({d}_j | \gamma, {\theta}_j)\, p({\theta}_j)
\end{equation}

\noindent
where the per-system parameters,
$${\theta}_j = \{a_j, e_j, i_j, \Omega_j, \omega_j, M_j, M_{1,j}, M_{2,j}\}$$  
and, the observables,
$${d}_j = \{r_{\perp,j}, \Delta v_{r,j}, \Delta\mu_{\alpha,j}, \Delta\mu_{\delta,j}\}$$ 

We adopt priors following \citetalias{saadting2025} with minor modifications. We place a uniform prior, $\Gamma \sim \mathcal{U}(-1, 1)$, where $\Gamma = \log_{10}\sqrt{\gamma}$ and $\gamma$ is the gravity boost factor. For masses, we use Gaussian priors centered on the \textit{Gaia} FLAME or \cite{Chae2023} estimates. For angular elements, we assign isotropic priors ($\cos i \sim \mathcal{U}(-1, 1)$; $\omega, \Omega, M \sim \mathcal{U}(0, 2\pi)$), and for eccentricity, we use the separation-dependent thermal distribution of \cite{Hwang2022eccentricity}, drawing $e \sim \mathrm{Beta}(\alpha_k+1,\,1)$ with a shape index $\alpha_k$ that depends on the projected-separation bin $k$.

The prior on the semi-major axis is important to this analysis. Following the observed sky-projected separation distribution of wide binaries \citep{Andrews2017, Pittordis2023}, we adopt a broken power-law prior on $a$:
\begin{equation}
P(a)\,\mathrm{d}a \propto \begin{cases} a^{-1}\,\mathrm{d}a & a \leq 5\,\mathrm{kAU} \\ a^{-1.6}\,\mathrm{d}a & a > 5\,\mathrm{kAU} \end{cases}
\label{eq:sma_prior}
\end{equation}
with a break at $a_{\rm break} = 5\,\mathrm{kAU}$ \citep{Pittordis2023}. Below the break the prior follows \"{O}pik's law; above the break it follows the power-law slope $\beta = -1.6$ measured by \citet{Andrews2017} from the observed wide-binary separation distribution. We sample $\log(a/r_{\perp}) \sim \mathcal{U}(-1.0,\,2.5)$ and apply a log-probability correction (PyMC \texttt{Potential}) to reweight from the base \"{O}pik distribution to the target broken power law, with continuity enforced at $a_{\rm break}$.

The likelihood follows \citetalias{saadting2025}. Each system contributes three terms:
\begin{align}
r_{\perp,\rm obs} &\sim \mathcal{N}(r_{\perp,\rm model},\; \sigma_{r}^2) \\
\Delta v_{r,\rm obs} &\sim t_5\!\left(\Delta v_{r,\rm model},\; \sigma_{\Delta v_r}^2 + \sigma_{\rm jitter}^2\right) \\
\Delta\mu_{\rm obs} &\sim \mathcal{N}(\Delta\mu_{\rm model},\; \sigma_{\mu}^2 + \sigma_{\rm pm,jit}^2)
\end{align}
where $t_5$ denotes a Student-$t$ distribution with five degrees of freedom, used for the RV likelihood to provide robustness against outliers. The parameters $\sigma_{\rm jitter}$ and $\sigma_{\rm pm,jit}$ are global nuisance parameters with half-normal priors. We perform inference using Hamiltonian Monte Carlo (HMC) with the No-U-Turn Sampler (NUTS) \citep{hoffman2014no}, implemented in PyMC \citep{pymc}, following the same approach as \citetalias{saadting2025}. We run 4 chains, each with 2000 tuning steps and 3000 sampling steps, with a target acceptance probability of 0.98. Convergence is assessed using the Gelman-Rubin statistic \citep{GelmanRubin1992}, with $\hat{R} < 1.01$ for $\gamma$ in all runs, and the bulk effective sample size (ESS) for $\gamma$ exceeding 400 in all cases.

\subsection{Using two different models} \label{sec:variants}

We consider two model variants that differ in how the three-dimensional separation is determined:

For the baseline model, the semi-major axis $a$ is a free parameter with the broken power-law prior of Equation~\ref{eq:sma_prior}, and the instantaneous separation $r_{\rm true}$ is computed from $a$ and $E$ via Equation~\ref{eq:rtrue}. The projected separation $r_\perp$ enters the likelihood as an observable with Gaussian uncertainty. This allows the model to infer an orbital scale independent of the observed projected separation.

We also consider a geometric de-projection model. In this variant, we replace the semi-major axis parameter with a geometric de-projection of the observed projected separation, following the approach of \cite{Chae2026}. The true separation is derived as,

\begin{equation}
    r_{\rm true} = \frac{r_\perp}{\sqrt{\cos^2\phi + \cos^2 i\, \sin^2\phi}}
\label{eq:rtrue_geom}
\end{equation} 

\noindent
where $\phi$ is the orbital position angle and $i$ is the inclination. This couples the three-dimensional orbital scale directly to the observed $r_\perp$ and does not allow for an independent determination of $a$. All other aspects of the model remain identical to the baseline.

\section{Results \& Implications} \label{sec:discussion}
\subsection{Results}

\begin{deluxetable}{lccc}
\tablecaption{Model Comparison Results\label{tab:tests}}
\tablehead{
\colhead{Model} & \colhead{$\gamma$} & \colhead{68\% CI} & \colhead{$P(\gamma < 1)$}
}
\startdata
Broken power-law prior & 1.00 & [0.81, 1.24] & 0.50 \\
Geometric $r_{\rm true}$ & 1.56 & [1.38, 1.77] & 0.00 \\
\enddata
\tablecomments{The broken power-law prior model includes an independent semi-major axis parameter with a broken power-law prior on $a$ following \citet{Andrews2017} and \citet{Pittordis2023} (Equation~\ref{eq:sma_prior}). The geometric $r_{\rm true}$ model replaces it with a de-projection of the observed projected separation, analogous to the approach of \cite{Chae2026}. The \cite{Chae2026} result is $\gamma \approx 1.60_{-0.14}^{+0.17}$. The median posterior $\gamma$, 68\% credible interval, and posterior probability of a gravity enhancement P($\gamma > 1$) are listed for each model.}
\end{deluxetable}

\begin{figure*}
    \centering
    \includegraphics[width=\columnwidth]{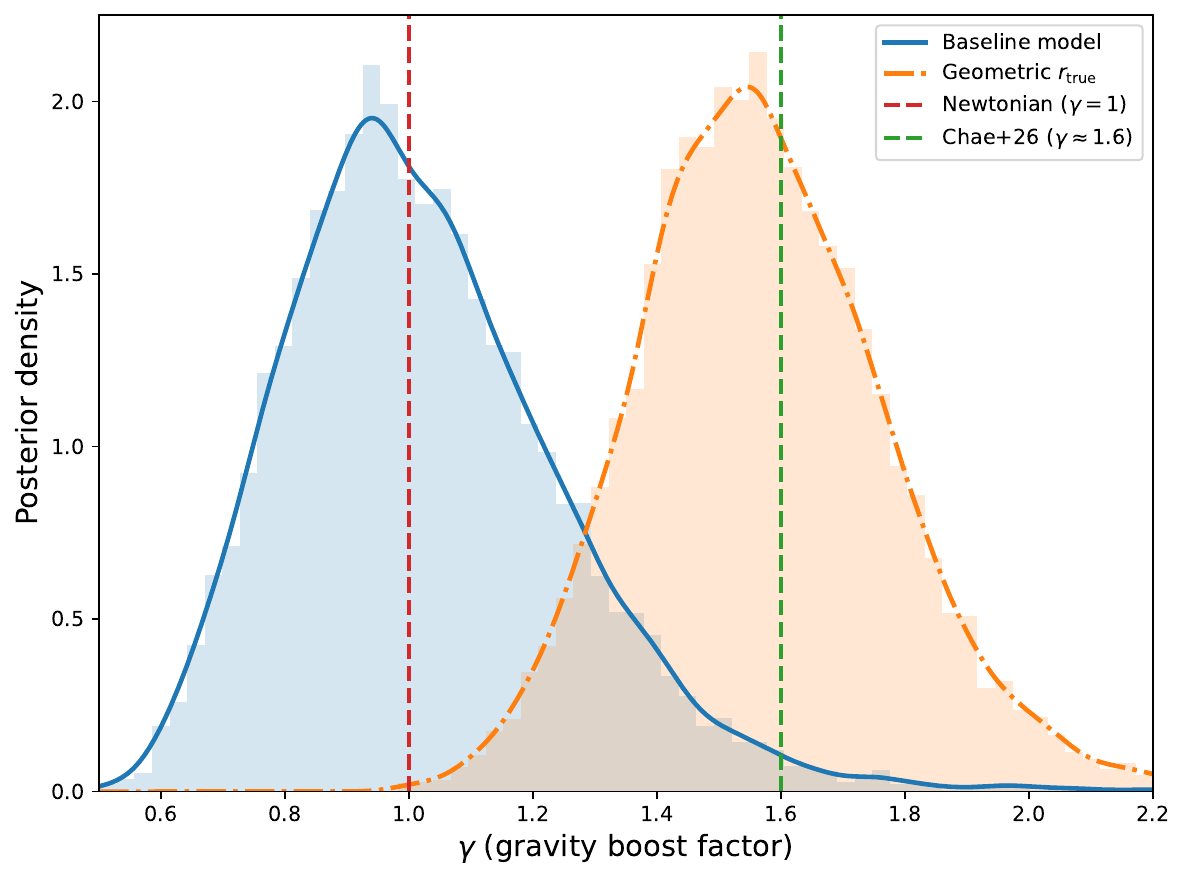}
    \includegraphics[width=\columnwidth]{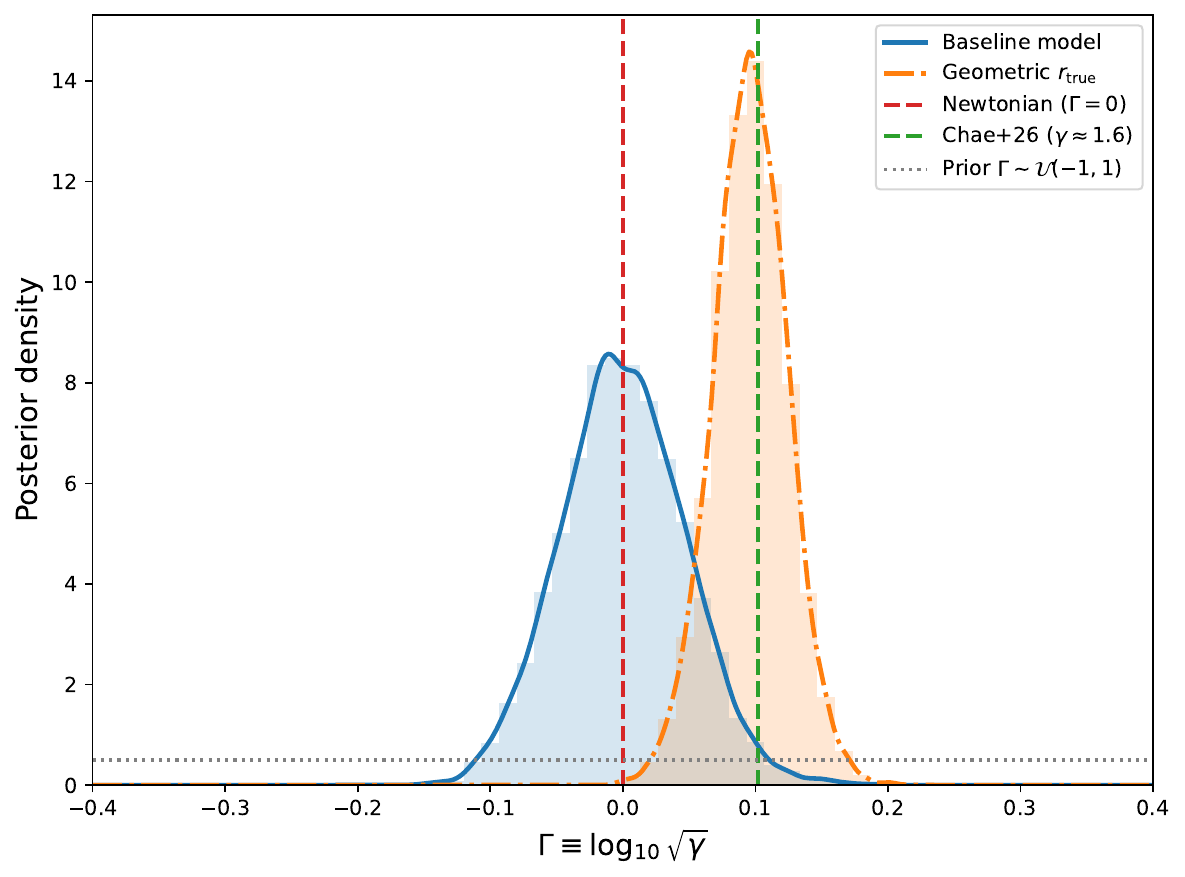}
    \caption{
    Posterior distributions for the gravitational boost parameter. 
    \emph{Left:} posterior distributions of $\gamma$ for the two model variants. 
    The blue solid curve shows the broken power-law prior model ($\gamma = 1.00$), which includes an independent semi-major axis parameter with a physically motivated broken power-law prior.
    The orange dashed curve shows the geometric de-projection model ($\gamma = 1.56$), in which $r_{\rm true}$ is derived directly from the observed projected separation. 
    The vertical red dashed line marks the Newtonian prediction ($\gamma = 1$), while the vertical green dashed line indicates $\gamma \approx 1.6$ reported by \citet{Chae2026}. 
    \emph{Right:} the same posteriors expressed in terms of $\Gamma \equiv \log_{10}\sqrt{\gamma}$. 
    The horizontal dotted line shows the flat prior $\Gamma \sim \mathcal{U}(-1,1)$ used in the inference. 
    The two models differ only in the treatment of the orbital separation, yet the inferred gravitational boost shifts substantially.
    }
    \label{fig:gamma_overlay}
\end{figure*}

\begin{figure*}
    \centering
    {\begin{tikzpicture}[
    node distance=0.85cm and 1.5cm,
    box/.style={draw, rounded corners=3pt, minimum width=2.6cm, minimum height=0.7cm, 
                align=center, font=\small},
    freeparam/.style={box, fill=blue!15, draw=black, thick},
    fixed/.style={box, fill=orange!15, draw=black, thick},
    shared/.style={box, fill=gray!12, draw=black},
    obs/.style={box, fill=yellow!20, draw=black},
    result/.style={box, fill=red!10, draw=black, thick},
    branch/.style={draw, circle, minimum size=0.5cm, fill=white, thick, inner sep=0pt, font=\footnotesize\bfseries},
    arrow/.style={-{Stealth[length=2.5mm]}, thick, black},
    bluearrow/.style={-{Stealth[length=2.5mm]}, thick, black},
    orangearrow/.style={-{Stealth[length=2.5mm]}, thick, black},
    label/.style={font=\footnotesize\itshape, text=black},
    pathlabel/.style={font=\small\bfseries, rounded corners=2pt, inner sep=3pt},
]

\node[freeparam] (gamma) at (0, 0) {$\gamma$, $M_1,\, M_2$, $e,\, i,\, \omega,\, \Omega,\, \phi$};

\coordinate (fork2) at (0, -1.5);
\node[label] at (0, -2) {How is $r_{\rm true}$ determined?};

\draw[arrow] (gamma) -- (fork2);

%

\node[freeparam] (sma) at (-3.8, -3.0) {$a$};
\node[shared, below=0.6cm of sma, minimum width=3.2cm] (kepler) {Baseline model:\\[-1pt]{\scriptsize $r_{\rm true} = a(1-e^2)/(1+e\cos\nu)$}};

\draw[arrow] (fork2) -| (sma);
\draw[arrow] (sma) -- (kepler);


\node[draw, rounded corners=5pt, dashed, thick,
      fit=(sma)(kepler), inner sep=6pt] {};

\node[obs] (rperp) at (3.8, -3.0) {$r_\perp$};
\node[shared, below=0.6cm of rperp, minimum width=3.4cm] (deproj) {De-projection:\\[-1pt]{\scriptsize $r_{\rm true} = r_\perp / \sqrt{\cos^2\!\phi + \cos^2\!i\,\sin^2\!\phi}$}};

\draw[arrow] (fork2) -| (rperp);
\draw[arrow] (rperp) -- (deproj);


\node[draw, rounded corners=5pt, dashed, thick,
      fit=(rperp)(deproj), inner sep=6pt] {};

\node[shared, thick, draw] (rtrue) at (0, -5.2) {$r_{\rm true}$};

\draw[arrow] (kepler) -| (rtrue);
\draw[arrow] (deproj) -| (rtrue);

\node[shared, below=0.7cm of rtrue] (velocity) {$v = \sqrt{\gamma\, G\, M_{\rm tot}\, F\, /\, r_{\rm true}}$};
\node[shared, below=0.7cm of velocity] (rotate) {Rotate to ICRS};
\node[obs, below=0.7cm of rotate] (observables) {Likelihood using:\\$r_\perp$, $\Delta v_r$, $\Delta\mu$};

\draw[arrow] (rtrue) -- (velocity);
\draw[arrow] (velocity) -- (rotate);
\draw[arrow] (rotate) -- (observables);

\node[result, minimum width=2.2cm] (res1) at (-3.8, -9.2) {$\gamma = 1.00_{-0.19}^{+0.24}$};
\node[result, minimum width=2.2cm] (res2) at (3.8, -9.2) {$\gamma = 1.56_{-0.18}^{+0.21}$};

\draw[arrow, dashed] (observables.south) ++(-0.3,0) -- ++(0, -0.4) -| (res1);
\draw[arrow, dashed] (observables.south) ++(0.3,0) -- ++(0, -0.4) -| (res2);


\node[draw, rounded corners=4pt, fill=white, align=left, anchor=north east] at (-6.5, 1) {
\begin{tikzpicture}[x=1cm,y=1cm]
\node[freeparam, minimum width=0.5cm, minimum height=0.3cm] at (0,0) {};
\node[anchor=west, font=\footnotesize] at (0.5,-0.15) {Free parameters};
\node[shared, minimum width=0.5cm, minimum height=0.3cm] at (0,-0.45) {};
\node[anchor=west, font=\footnotesize] at (0.5,-0.6) {Intermediate steps};
\node[obs, minimum width=0.5cm, minimum height=0.3cm] at (0,-0.9) {};
\node[anchor=west, font=\footnotesize] at (0.5,-1.05) {Observed parameters};
\node[result, minimum width=0.5cm, minimum height=0.3cm] at (0,-1.35) {};
\node[anchor=west, font=\footnotesize] at (0.5,-1.5) {Results};
\end{tikzpicture}
};

\end{tikzpicture}}
\caption{Diagram showing how the two model variants diverge in their determination of the three-dimensional separation $r_{\rm true}$, while sharing all other components. At the branch point, broken power-law prior model treats the semi-major axis $a$ as a free parameter with a broken power-law prior and computes $r_{\rm true}$ from Kepler's equation, yielding $\gamma = 1.00_{-0.19}^{+0.24}$, consistent with Newtonian gravity. Geometric model derives $r_{\rm true}$ directly from the observed projected separation $r_\perp$ through geometric de-projection, yielding $\gamma = 1.56_{-0.18}^{+0.21}$, consistent with the anomaly reported by \citet{Chae2026}. All other steps, like velocity computation, coordinate rotation, and comparison to observables are identical between the two models.}
\label{fig:schematic}
\end{figure*}

The results of both model variants are summarized in Table~\ref{tab:tests} and Figure~\ref{fig:gamma_overlay}. For the broken power-law prior model, our hierarchical Bayesian analysis yields $\gamma = 1.00$ with a 68\% credible interval of $[0.81, 1.24]$ and a 95\% credible interval of $[0.66, 1.54]$. The Newtonian value $\gamma = 1$ is at the centre of the posterior. The probability that $\gamma > 1$ is 0.50. In the notation of \cite{Chae2026}, this corresponds to $\Gamma = \log_{10}\sqrt{\gamma} \approx 0.00$.

Replacing the semi-major axis with the geometric de-projection shifts the posterior to $\gamma = 1.56$ with a 68\% credible interval of $[1.38, 1.77]$ and a 95\% credible interval of $[1.20, 2.01]$. The probability that $\gamma > 1$ is effectively unity. This result is consistent with the \cite{Chae2026} value of $\gamma \approx 1.60_{-0.14}^{+0.17}$.

The two models are identical in every respect except the treatment of the three-dimensional separation, yet the inferred $\gamma$ differs by $\Delta\gamma \approx 0.56$, shifting from a result consistent with Newtonian gravity to one consistent with the reported gravitational anomaly. Figure~\ref{fig:schematic} summarizes the point at which the two models diverge and how that difference propagates to the inferred value of $\gamma$.

\subsection{The role of the semi-major axis}

In the two different models we used, we parametrize the orbital scale differently. In the baseline model, $a$ is treated as an independent parameter. The instantaneous separation is then computed from,
\begin{equation}
    r_{\rm true} = a\,(1 - e\cos E)
\end{equation}
and the projected separation is obtained by projecting the orbit onto the plane of the sky. The observed $r_\perp$ enters through the likelihood, so the model asks whether a given orbit predicts a projected separation consistent with the data. In the geometric de-projection model, the observed projected separation is used to define the instantaneous three-dimensional separation (Equation \ref{eq:rtrue_geom}). The semi-major axis is not used in this parametrization, but it can be obtained from,
\begin{equation}
    a = \frac{r_{\rm true}}{1 - e\cos E}
\end{equation}
In this case, $a$ is not an independent parameter. The orbital scale is tied directly to the observed projected separation and the orbital geometry.


The two models therefore use the same Keplerian relations but place the projected separation at different points in the inference. In the baseline model, $r_\perp$ is a predicted observable. In the geometric model, $r_\perp$ is used to define part of the orbit itself. This changes the set of allowed combinations of $r_{\rm true}$, $a$, and $\gamma$.

This difference matters because, through the vis-viva relation, the velocity (Equation~\ref{eq:vmag}) depends on \emph{both} the instantaneous separation $r_{\rm true}$ and the semi-major axis $a$. In the geometric model, once $r_\perp$, $i$, and the orbital phase are specified, $r_{\rm true}$ is fixed by de-projection and $a$ then follows from it, so the orbital scale is tied directly to the observable. In the baseline model, $a$ is an independent parameter and $r_{\rm true}$ is computed from it via Equation~\ref{eq:rtrue}, so the prior on $a$ determines the range of $r_{\rm true}$ values the sampler can explore. The same observed system can therefore be assigned different posterior support in the two parameterizations.

In our analysis, this leads to different inferences for $\gamma$. When $a$ is treated as an independent parameter with a broken power-law prior, we obtain $\gamma = 1.00_{-0.19}^{+0.24}$, which is consistent with Newtonian gravity. When the orbital scale is set by geometric de-projection of the observed projected separation, we obtain $\gamma = 1.56_{-0.18}^{+0.21}$, consistent with \cite{Chae2026}. This comparison shows that for the present dataset, the inferred value of $\gamma$ is sensitive to how the orbital scale is introduced into the model.

\subsection{Additional methodological differences}

Beyond the semi-major axis treatment, several other differences exist between our approach and that of \cite{Chae2026} that may contribute at a smaller level.

Firstly, we use NUTS/HMC in \texttt{PyMC}, which explores high-dimensional correlated posteriors efficiently, while \cite{Chae2026} uses \texttt{emcee} with 200 walkers. For a six-parameter per-system model, \texttt{emcee} is adequate, and this difference is unlikely to drive the discrepancy in the inferred $\gamma$.

Secondly, we use a hierarchical model where $\gamma$ is a global parameter fit jointly across all systems, whereas \cite{Chae2026} fits $\Gamma$ for each system independently and consolidates the per-system posteriors using the method of \cite{HillMiller2011}. Both approaches weight systems by how well their orbits are constrained: in a hierarchical framework this happens automatically through the joint likelihood, while in the \cite{Chae2026} approach the consolidation step accounts for per-system posterior widths. This difference is also unlikely to be the primary driver of the discrepancy.

Thirdly, the treatment of the projected separation may also play a role. In our baseline model, $r_\perp$ enters as an observable with Gaussian uncertainty, so that any error in $r_\perp$ is propagated into the posterior on $\gamma$. In the \cite{Chae2026} model, $r_\perp$ is treated as a fixed quantity. Since $v^2 \propto \gamma / r$, an underestimated separation could contribute to an overestimated $\gamma$. To test whether this treatment matters in practice, we run an additional variant of the geometric de-projection model in which $r_\perp$ is treated as exact and the $r_\perp$ likelihood term is removed. This yields $\gamma = 1.59^{+0.23}_{-0.19}$, nearly identical to the geometric model with $r_\perp$ likelihood included ($\gamma = 1.56^{+0.21}_{-0.18}$). The shift of $\Delta\gamma \approx 0.03$ confirms that the $r_\perp$ treatment is subdominant compared to the semi-major axis parameterization. We present this comparison in detail in Appendix~\ref{appendix:rproj_test}.

As a check on the inferred orbital parameters, we present the population-level posterior distribution of the true anomaly in Appendix~\ref{appendix:nu}. As an independent validation of the inference procedure itself, we perform an injection-recovery test in Appendix~\ref{appendix:injection}: we generate synthetic catalogs with a known input gravity boost factor and find that the baseline model recovers it without bias. A small number of systems in this sample have relative velocities above the largest Newtonian escape velocity allowed by their measured masses and projected separations. We discuss these systems, the possible causes, and the sense in which our result is a population-level statement, in Appendix~\ref{appendix:highv}.

\subsection{Relation to previous work}

Our result is consistent with our previous analysis \citep{saadting2025}, where we applied a similar hierarchical Bayesian framework to the C3PO wide-binary sample and found the canonical MOND value of $a_0 = 1.2 \times 10^{-10}~\rm{m\,s^{-2}}$ to be excluded at $\sim3\sigma$ and $\sim2\sigma$ for two different interpolating functions. The present analysis is more informative because it uses the same data as \cite{Chae2026}, who reported a gravitational anomaly in these 36 systems. It also differs from our previous work in that here we estimate $\gamma$, which is independent of the choice of interpolating function or EFE treatment.

Our result is also consistent with the findings of \cite{Banik2024} and \cite{Pittordis2023}, with the latter updated in \cite{Pittordis2025}, who used ensemble statistics of \textit{Gaia} proper motions to conclude that Newtonian gravity is preferred. We note that \cite{Saglia2025}, who provided many of the HARPS RVs used by \cite{Chae2026}, found that Newtonian Keplerian orbits provide adequate fits for most of their systems. Their analysis did not fit for a gravity boost parameter; rather, it demonstrated that the observed kinematics are consistent with Newtonian orbits. Because our model reduces to Newtonian Keplerian motion at $\gamma=1$, our inferred orbital parameters in this limit are, by construction, those of a Newtonian orbit fit, and are consistent with the Newtonian solutions of \cite{Saglia2025} for the overlapping systems. The fact that \cite{Chae2026} infers $\gamma >1$ from overlapping data underscores that the inferred anomaly depends on the modeling framework, consistent with the sensitivity to orbital parameterization that we identify in this work.

\section{Conclusions} \label{sec:conclusions}

We reanalyzed the 36 wide binaries from \citet{Chae2026} using a hierarchical Bayesian model that fits a global gravity boost factor $\gamma$ while forward modelling three-dimensional Keplerian orbits for each system. In our baseline model, where the semi-major axis is treated as an independent parameter with a broken power-law prior $P(a) \propto a^{-1.6}$ for $a > 5\,\mathrm{kAU}$ \citep{Andrews2017, Pittordis2023}, we find $\gamma = 1.00^{+0.24}_{-0.19}$, consistent with Newtonian gravity. This differs from the result of \citet{Chae2026}, who reported $\gamma \approx 1.60_{-0.14}^{+0.17}$ for the same systems.

To understand this difference, we repeated the analysis with a second model that follows the geometric de-projection approach used by \citet{Chae2026}, where the three-dimensional separation is derived directly from the observed projected separation and no independent semi-major axis is included. This model gives $\gamma = 1.56^{+0.21}_{-0.18}$, in close agreement with \citet{Chae2026}. This comparison indicates that the inferred value of $\gamma$ depends on how the orbital scale is introduced into the inference. The forward modelling approach, in which the semi-major axis is a free parameter and the projected separation is predicted and compared to the data, is the more rigorous method, as it avoids inferring latent 3D variables directly from observables. The geometric de-projection approach of \citet{Chae2026} uses the projected separation to define the orbit, which is less rigorous when a wide range of 3D configurations can produce the same observable. Our result demonstrates that the wide binaries in this sample are consistent with Newtonian gravity when analyzed with a physically motivated prior on the semi-major axis and a proper forward modelling approach. The claimed gravitational anomaly of \citet{Chae2026} is not supported by our analysis.

\section*{Acknowledgments}

We thank Indranil Banik, Xavier Hernandez and Kyu-Hyun Chae for helpful discussions.

SMS is supported by the Distinguished University Fellowship awarded by The Ohio State University. YST is supported by NSF under Grant AST-2406729 and a Humboldt Research Award from the Alexander von Humboldt Foundation.

This work has made use of data from the European Space Agency (ESA) mission \textit{Gaia} (\url{https://www.cosmos.esa.int/gaia}), processed by the \textit{Gaia} Data Processing and Analysis Consortium (DPAC).

This work has made use of the following software packages: AstroPy \citep{astropy}, PyMC \citep{pymc}, and ArviZ \citep{arviz}. This project has also benefited from the use of Cursor and Claude Code. These tools were used for code development, for generating the figures, and for drafting and editing portions of the text. The authors designed the analysis, verified all numerical results, and take full responsibility for the content of this paper.

\clearpage
\section*{Data \& Code Availability}
The data underlying this article are available in the published tables and supplementary materials of \citet{Chae2026}. The analysis code and derived posterior samples are available on \url{https://github.com/seratsaad/wb3d-gamma}. The details of the analysis done by \cite{Chae2026} can be found in Zenodo at \cite{Chae2025soft}.

\bibliographystyle{mnras}
\bibliography{references}

\begin{thebibliography}{}
\makeatletter
\relax
\def\mn@urlcharsother{\let\do\@makeother \do\$\do\&\do\#\do\^\do\_\do\%\do\~}
\def\mn@doi{\begingroup\mn@urlcharsother \@ifnextchar [ {\mn@doi@}
  {\mn@doi@[]}}
\def\mn@doi@[#1]#2{\def\@tempa{#1}\ifx\@tempa\@empty \href
  {http://dx.doi.org/#2} {doi:#2}\else \href {http://dx.doi.org/#2} {#1}\fi
  \endgroup}
\def\mn@eprint#1#2{\mn@eprint@#1:#2::\@nil}
\def\mn@eprint@arXiv#1{\href {http://arxiv.org/abs/#1} {{\tt arXiv:#1}}}
\def\mn@eprint@dblp#1{\href {http://dblp.uni-trier.de/rec/bibtex/#1.xml}
  {dblp:#1}}
\def\mn@eprint@#1:#2:#3:#4\@nil{\def\@tempa {#1}\def\@tempb {#2}\def\@tempc
  {#3}\ifx \@tempc \@empty \let \@tempc \@tempb \let \@tempb \@tempa \fi \ifx
  \@tempb \@empty \def\@tempb {arXiv}\fi \@ifundefined
  {mn@eprint@\@tempb}{\@tempb:\@tempc}{\expandafter \expandafter \csname
  mn@eprint@\@tempb\endcsname \expandafter{\@tempc}}}

\bibitem[\protect\citeauthoryear{{Abdurro'uf} et~al.,}{{Abdurro'uf}
  et~al.}{2022}]{sdssdr17apogee}
{Abdurro'uf} et~al., 2022, \mn@doi [\apjs] {10.3847/1538-4365/ac4414}, \href
  {https://ui.adsabs.harvard.edu/abs/2022ApJS..259...35A} {259, 35}

\bibitem[\protect\citeauthoryear{{Andrews}, {Chanam{\'e}}  \&
  {Ag{\"u}eros}}{{Andrews} et~al.}{2017}]{Andrews2017}
{Andrews} J.~J.,  {Chanam{\'e}} J.,   {Ag{\"u}eros} M.~A.,  2017, \mn@doi
  [\mnras] {10.1093/mnras/stx2000}, \href
  {https://ui.adsabs.harvard.edu/abs/2017MNRAS.472..675A} {472, 675}

\bibitem[\protect\citeauthoryear{{Angus}}{{Angus}}{2008}]{Angus2008}
{Angus} G.~W.,  2008, \mn@doi [\mnras] {10.1111/j.1365-2966.2008.13351.x},
  \href {https://ui.adsabs.harvard.edu/abs/2008MNRAS.387.1481A} {387, 1481}

\bibitem[\protect\citeauthoryear{{Asencio}, {Banik}, {Mieske}, {Venhola},
  {Kroupa}  \& {Zhao}}{{Asencio} et~al.}{2022}]{Asencio2022}
{Asencio} E.,  {Banik} I.,  {Mieske} S.,  {Venhola} A.,  {Kroupa} P.,   {Zhao}
  H.,  2022, \mn@doi [\mnras] {10.1093/mnras/stac1765}, 515, 2981

\bibitem[\protect\citeauthoryear{{Astropy Collaboration} et~al.,}{{Astropy
  Collaboration} et~al.}{2013}]{astropy}
{Astropy Collaboration} et~al., 2013, \mn@doi [\aap]
  {10.1051/0004-6361/201322068}, \href
  {https://ui.adsabs.harvard.edu/abs/2013A&A...558A..33A} {558, A33}

\bibitem[\protect\citeauthoryear{{Banik} \& {Zhao}}{{Banik} \&
  {Zhao}}{2018}]{Banik2018}
{Banik} I.,  {Zhao} H.,  2018, \mn@doi [\mnras] {10.1093/mnras/sty2007}, \href
  {https://ui.adsabs.harvard.edu/abs/2018MNRAS.480.2660B} {480, 2660}

\bibitem[\protect\citeauthoryear{{Banik} \& {Zhao}}{{Banik} \&
  {Zhao}}{2022}]{BanikZhao2022}
{Banik} I.,  {Zhao} H.,  2022, \mn@doi [Symmetry] {10.3390/sym14071331}, \href
  {https://ui.adsabs.harvard.edu/abs/2022Symm...14.1331B} {14, 1331}

\bibitem[\protect\citeauthoryear{{Banik}, {Pittordis}, {Sutherland}, {Famaey},
  {Ibata}, {Mieske}  \& {Zhao}}{{Banik} et~al.}{2024}]{Banik2024}
{Banik} I.,  {Pittordis} C.,  {Sutherland} W.,  {Famaey} B.,  {Ibata} R.,
  {Mieske} S.,   {Zhao} H.,  2024, \mn@doi [\mnras] {10.1093/mnras/stad3393},
  \href {https://ui.adsabs.harvard.edu/abs/2024MNRAS.527.4573B} {527, 4573}

\bibitem[\protect\citeauthoryear{{Chae}}{{Chae}}{2023}]{Chae2023}
{Chae} K.-H.,  2023, \mn@doi [\apj] {10.3847/1538-4357/ace101}, \href
  {https://ui.adsabs.harvard.edu/abs/2023ApJ...952..128C} {952, 128}

\bibitem[\protect\citeauthoryear{{Chae}}{{Chae}}{2024}]{Chae2024a}
{Chae} K.-H.,  2024, \mn@doi [\apj] {10.3847/1538-4357/ad61e9}, \href
  {https://ui.adsabs.harvard.edu/abs/2024ApJ...972..186C} {972, 186}

\bibitem[\protect\citeauthoryear{Chae}{Chae}{2025a}]{Chae2025soft}
Chae K.-H.,  2025a, Bayesian 3D modeling of the orbital dynamics of wide binary
  stars: General Python algorithms to infer gravity,
  \mn@doi{10.5281/zenodo.17113129}

\bibitem[\protect\citeauthoryear{{Chae}}{{Chae}}{2025b}]{Chae2025harps}
{Chae} K.-H.,  2025b, \mn@doi [arXiv e-prints] {10.48550/arXiv.2508.11996},
  \href {https://ui.adsabs.harvard.edu/abs/2025arXiv250811996C} {p.
  arXiv:2508.11996}

\bibitem[\protect\citeauthoryear{{Chae} \& {Yoon}}{{Chae} \&
  {Yoon}}{2026}]{ChaeYoon2026}
{Chae} K.-H.,  {Yoon} Y.,  2026, arXiv e-prints

\bibitem[\protect\citeauthoryear{{Chae}, {Lelli}, {Desmond}, {McGaugh}  \&
  {Schombert}}{{Chae} et~al.}{2022}]{Chae2022ext}
{Chae} K.-H.,  {Lelli} F.,  {Desmond} H.,  {McGaugh} S.~S.,   {Schombert}
  J.~M.,  2022, \mn@doi [\prd] {10.1103/PhysRevD.106.103025}, \href
  {https://ui.adsabs.harvard.edu/abs/2022PhRvD.106j3025C} {106, 103025}

\bibitem[\protect\citeauthoryear{{Chae}, {Lee}, {Hernandez}, {Orlov}, {Lim},
  {Turnshek}  \& {Lee}}{{Chae} et~al.}{2026}]{Chae2026}
{Chae} K.-H.,  {Lee} B.-C.,  {Hernandez} X.,  {Orlov} V.~G.,  {Lim} D.,
  {Turnshek} D.~A.,   {Lee} Y.-W.,  2026, \mn@doi [arXiv e-prints]
  {10.48550/arXiv.2601.21728}, \href
  {https://ui.adsabs.harvard.edu/abs/2026arXiv260121728C} {p. arXiv:2601.21728}

\bibitem[\protect\citeauthoryear{{Cookson}, {Banik}, {El-Badry}, {Sutherland},
  {Penoyre}, {Pittordis}  \& {Clarke}}{{Cookson} et~al.}{2026}]{Cookson2026}
{Cookson} S.~A.,  {Banik} I.,  {El-Badry} K.,  {Sutherland} W.,  {Penoyre} Z.,
  {Pittordis} C.,   {Clarke} C.~J.,  2026, \mn@doi [\mnras]
  {10.1093/mnras/stag342}, \href
  {https://ui.adsabs.harvard.edu/abs/2026MNRAS.547ag342C} {547, stag342}

\bibitem[\protect\citeauthoryear{{Desmond}, {Hees}  \& {Famaey}}{{Desmond}
  et~al.}{2024}]{Desmond2024}
{Desmond} H.,  {Hees} A.,   {Famaey} B.,  2024, \mn@doi [\mnras]
  {10.1093/mnras/stae955}, \href
  {https://ui.adsabs.harvard.edu/abs/2024MNRAS.530.1781D} {530, 1781}

\bibitem[\protect\citeauthoryear{{Eisenstein} et~al.,}{{Eisenstein}
  et~al.}{2005}]{Eisenstein2005}
{Eisenstein} D.~J.,  et~al., 2005, \mn@doi [\apj] {10.1086/466512}, \href
  {https://ui.adsabs.harvard.edu/abs/2005ApJ...633..560E} {633, 560}

\bibitem[\protect\citeauthoryear{{Faber} \& {Gallagher}}{{Faber} \&
  {Gallagher}}{1979}]{Faber1979}
{Faber} S.~M.,  {Gallagher} J.~S.,  1979, \mn@doi [\araa]
  {10.1146/annurev.aa.17.090179.001031}, \href
  {https://ui.adsabs.harvard.edu/abs/1979ARA&A..17..135F} {17, 135}

\bibitem[\protect\citeauthoryear{{Famaey} \& {McGaugh}}{{Famaey} \&
  {McGaugh}}{2012}]{Famaey2012}
{Famaey} B.,  {McGaugh} S.~S.,  2012, \mn@doi [Living Reviews in Relativity]
  {10.12942/lrr-2012-10}, \href
  {https://ui.adsabs.harvard.edu/abs/2012LRR....15...10F} {15, 10}

\bibitem[\protect\citeauthoryear{{Gelman} \& {Rubin}}{{Gelman} \&
  {Rubin}}{1992}]{GelmanRubin1992}
{Gelman} A.,  {Rubin} D.~B.,  1992, \mn@doi [Statistical Science]
  {10.1214/ss/1177011136}, \href
  {https://ui.adsabs.harvard.edu/abs/1992StaSc...7..457G} {7, 457}

\bibitem[\protect\citeauthoryear{{Hees}, {Folkner}, {Jacobson}  \&
  {Park}}{{Hees} et~al.}{2014}]{Hees2014}
{Hees} A.,  {Folkner} W.~M.,  {Jacobson} R.~A.,   {Park} R.~S.,  2014, \mn@doi
  [\prd] {10.1103/PhysRevD.89.102002}, 89, 102002

\bibitem[\protect\citeauthoryear{{Hernandez}, {Verteletskyi}, {Nasser}  \&
  {Aguayo-Ortiz}}{{Hernandez} et~al.}{2024}]{Hernandez2024}
{Hernandez} X.,  {Verteletskyi} V.,  {Nasser} L.,   {Aguayo-Ortiz} A.,  2024,
  \mn@doi [\mnras] {10.1093/mnras/stad3446}, \href
  {https://ui.adsabs.harvard.edu/abs/2024MNRAS.528.4720H} {528, 4720}

\bibitem[\protect\citeauthoryear{{Hill} \& {Miller}}{{Hill} \&
  {Miller}}{2011}]{HillMiller2011}
{Hill} T.~P.,  {Miller} J.,  2011, \mn@doi [Chaos] {10.1063/1.3593373}, \href
  {https://ui.adsabs.harvard.edu/abs/2011Chaos..21c3102H} {21, 033102}

\bibitem[\protect\citeauthoryear{Hoffman, Gelman  et~al.}{Hoffman
  et~al.}{2014}]{hoffman2014no}
Hoffman M.~D.,  Gelman A.,   et~al., 2014, J. Mach. Learn. Res., 15, 1593

\bibitem[\protect\citeauthoryear{{Hwang}, {Ting}  \& {Zakamska}}{{Hwang}
  et~al.}{2022}]{Hwang2022eccentricity}
{Hwang} H.-C.,  {Ting} Y.-S.,   {Zakamska} N.~L.,  2022, \mn@doi [\mnras]
  {10.1093/mnras/stac675}, \href
  {https://ui.adsabs.harvard.edu/abs/2022MNRAS.512.3383H} {512, 3383}

\bibitem[\protect\citeauthoryear{{Kahn} \& {Woltjer}}{{Kahn} \&
  {Woltjer}}{1959}]{Kahn1959}
{Kahn} F.~D.,  {Woltjer} L.,  1959, \mn@doi [\apj] {10.1086/146762}, \href
  {https://ui.adsabs.harvard.edu/abs/1959ApJ...130..705K} {130, 705}

\bibitem[\protect\citeauthoryear{{Kumar}, {Carroll}, {Hartikainen}  \&
  {Martin}}{{Kumar} et~al.}{2019}]{arviz}
{Kumar} R.,  {Carroll} C.,  {Hartikainen} A.,   {Martin} O.,  2019, \mn@doi
  [The Journal of Open Source Software] {10.21105/joss.01143}, \href
  {https://ui.adsabs.harvard.edu/abs/2019JOSS....4.1143K} {4, 1143}

\bibitem[\protect\citeauthoryear{{Lelli}, {McGaugh}, {Schombert}  \&
  {Pawlowski}}{{Lelli} et~al.}{2016}]{Lelli2016}
{Lelli} F.,  {McGaugh} S.~S.,  {Schombert} J.~M.,   {Pawlowski} M.~S.,  2016,
  \mn@doi [\apjl] {10.3847/2041-8205/827/1/L19}, \href
  {https://ui.adsabs.harvard.edu/abs/2016ApJ...827L..19L} {827, L19}

\bibitem[\protect\citeauthoryear{{Lelli}, {McGaugh}  \& {Schombert}}{{Lelli}
  et~al.}{2017}]{Lelli2017}
{Lelli} F.,  {McGaugh} S.~S.,   {Schombert} J.~M.,  2017, \mn@doi [\mnras]
  {10.1093/mnrasl/slx031}, \href
  {https://ui.adsabs.harvard.edu/abs/2017MNRAS.468L..68L} {468, L68}

\bibitem[\protect\citeauthoryear{{Mahmud Saad} \& {Ting}}{{Mahmud Saad} \&
  {Ting}}{2025}]{saadting2025}
{Mahmud Saad} S.,  {Ting} Y.-S.,  2025, \mn@doi [arXiv e-prints]
  {10.48550/arXiv.2512.19652}, \href
  {https://ui.adsabs.harvard.edu/abs/2025arXiv251219652M} {p. arXiv:2512.19652}

\bibitem[\protect\citeauthoryear{{McGaugh}}{{McGaugh}}{2016}]{Mcgaugh2016}
{McGaugh} S.~S.,  2016, \mn@doi [\apjl] {10.3847/2041-8205/832/1/L8}, \href
  {https://ui.adsabs.harvard.edu/abs/2016ApJ...832L...8M} {832, L8}

\bibitem[\protect\citeauthoryear{{Milgrom}}{{Milgrom}}{1983}]{Milgrom1983}
{Milgrom} M.,  1983, \mn@doi [\apj] {10.1086/161131}, \href
  {https://ui.adsabs.harvard.edu/abs/1983ApJ...270..371M} {270, 371}

\bibitem[\protect\citeauthoryear{{Ostriker} \& {Peebles}}{{Ostriker} \&
  {Peebles}}{1973}]{Ostriker1973}
{Ostriker} J.~P.,  {Peebles} P.~J.~E.,  1973, \mn@doi [\apj] {10.1086/152513},
  \href {https://ui.adsabs.harvard.edu/abs/1973ApJ...186..467O} {186, 467}

\bibitem[\protect\citeauthoryear{{Park}, {Hees}, {Famaey}, {Desmond}  \&
  {Durakovic}}{{Park} et~al.}{2026}]{Park2026}
{Park} R.~S.,  {Hees} A.,  {Famaey} B.,  {Desmond} H.,   {Durakovic} A.,  2026,
  \prd

\bibitem[\protect\citeauthoryear{{Pittordis} \& {Sutherland}}{{Pittordis} \&
  {Sutherland}}{2023}]{Pittordis2023}
{Pittordis} C.,  {Sutherland} W.,  2023, \mn@doi [The Open Journal of
  Astrophysics] {10.21105/astro.2205.02846}, \href
  {https://ui.adsabs.harvard.edu/abs/2023OJAp....6E...4P} {6, 4}

\bibitem[\protect\citeauthoryear{{Pittordis}, {Sutherland}  \&
  {Shepherd}}{{Pittordis} et~al.}{2025}]{Pittordis2025}
{Pittordis} C.,  {Sutherland} W.,   {Shepherd} P.,  2025, \mn@doi [The Open
  Journal of Astrophysics] {10.33232/001c.142887}, \href
  {https://ui.adsabs.harvard.edu/abs/2025OJAp....8E.109P} {8, 109}

\bibitem[\protect\citeauthoryear{{Planck Collaboration} et~al.,}{{Planck
  Collaboration} et~al.}{2020a}]{Planck2020a}
{Planck Collaboration} et~al., 2020a, \mn@doi [\aap]
  {10.1051/0004-6361/201833880}, \href
  {https://ui.adsabs.harvard.edu/abs/2020A&A...641A...1P} {641, A1}

\bibitem[\protect\citeauthoryear{{Planck Collaboration} et~al.,}{{Planck
  Collaboration} et~al.}{2020b}]{Planck2020b}
{Planck Collaboration} et~al., 2020b, \mn@doi [\aap]
  {10.1051/0004-6361/201833910}, \href
  {https://ui.adsabs.harvard.edu/abs/2020A&A...641A...6P} {641, A6}

\bibitem[\protect\citeauthoryear{{Rubin} \& {Ford}}{{Rubin} \&
  {Ford}}{1970}]{Rubin1970}
{Rubin} V.~C.,  {Ford} Jr. W.~K.,  1970, \mn@doi [\apj] {10.1086/150317}, \href
  {https://ui.adsabs.harvard.edu/abs/1970ApJ...159..379R} {159, 379}

\bibitem[\protect\citeauthoryear{{Russell}, {Banik}, {Cray}  \&
  {Zhao}}{{Russell} et~al.}{2026}]{Russell2026}
{Russell} A.,  {Banik} I.,  {Cray} O.,   {Zhao} H.,  2026, \mn@doi [\mnras]
  {10.1093/mnras/stag399}, 547

\bibitem[\protect\citeauthoryear{{Sacco} et~al.,}{{Sacco}
  et~al.}{2014}]{Gaiaflame}
{Sacco} G.~G.,  et~al., 2014, \mn@doi [\aap] {10.1051/0004-6361/201423619},
  \href {https://ui.adsabs.harvard.edu/abs/2014A&A...565A.113S} {565, A113}

\bibitem[\protect\citeauthoryear{{Saglia}, {Pasquini}, {Patat}, {Ludwig},
  {Giribaldi}, {Leao}, {de Medeiros}  \& {Murphy}}{{Saglia}
  et~al.}{2025}]{Saglia2025}
{Saglia} R.,  {Pasquini} L.,  {Patat} F.,  {Ludwig} H.-G.,  {Giribaldi} R.,
  {Leao} I.,  {de Medeiros} J.~R.,   {Murphy} M.~T.,  2025, \mn@doi [\aap]
  {10.1051/0004-6361/202555115}, \href
  {https://ui.adsabs.harvard.edu/abs/2025A&A...699A.151S} {699, A151}

\bibitem[\protect\citeauthoryear{{Salvatier}, {Wiecki}  \&
  {Fonnesbeck}}{{Salvatier} et~al.}{2015}]{pymc}
{Salvatier} J.,  {Wiecki} T.,   {Fonnesbeck} C.,  2015, \mn@doi [arXiv
  e-prints] {10.48550/arXiv.1507.08050}, \href
  {https://ui.adsabs.harvard.edu/abs/2015arXiv150708050S} {p. arXiv:1507.08050}

\bibitem[\protect\citeauthoryear{{Sanders}}{{Sanders}}{2003}]{Sanders2003}
{Sanders} R.~H.,  2003, \mn@doi [\mnras] {10.1046/j.1365-8711.2003.06596.x},
  \href {https://ui.adsabs.harvard.edu/abs/2003MNRAS.342..901S} {342, 901}

\bibitem[\protect\citeauthoryear{{Sanders} \& {McGaugh}}{{Sanders} \&
  {McGaugh}}{2002}]{Sanders2002}
{Sanders} R.~H.,  {McGaugh} S.~S.,  2002, \mn@doi [\araa]
  {10.1146/annurev.astro.40.060401.093923}, \href
  {https://ui.adsabs.harvard.edu/abs/2002ARA&A..40..263S} {40, 263}

\bibitem[\protect\citeauthoryear{{Sofue} \& {Rubin}}{{Sofue} \&
  {Rubin}}{2001}]{Sofue2001}
{Sofue} Y.,  {Rubin} V.,  2001, \mn@doi [\araa]
  {10.1146/annurev.astro.39.1.137}, \href
  {https://ui.adsabs.harvard.edu/abs/2001ARA&A..39..137S} {39, 137}

\bibitem[\protect\citeauthoryear{{Springel} et~al.,}{{Springel}
  et~al.}{2005}]{Springel2005}
{Springel} V.,  et~al., 2005, \mn@doi [\nat] {10.1038/nature03597}, \href
  {https://ui.adsabs.harvard.edu/abs/2005Natur.435..629S} {435, 629}

\bibitem[\protect\citeauthoryear{{Vokrouhlick{\'y}}, {Nesvorn{\'y}}  \&
  {Tremaine}}{{Vokrouhlick{\'y}} et~al.}{2024}]{Vokrouhlicky2024}
{Vokrouhlick{\'y}} D.,  {Nesvorn{\'y}} D.,   {Tremaine} S.,  2024, \mn@doi
  [\apj] {10.3847/1538-4357/ad40a3}, 968, 47

\bibitem[\protect\citeauthoryear{{Yong} et~al.,}{{Yong} et~al.}{2023}]{C3POI}
{Yong} D.,  et~al., 2023, \mn@doi [\mnras] {10.1093/mnras/stad2679}, \href
  {https://ui.adsabs.harvard.edu/abs/2023MNRAS.526.2181Y} {526, 2181}

\bibitem[\protect\citeauthoryear{{Zwicky}}{{Zwicky}}{1937}]{Zwicky1937}
{Zwicky} F.,  1937, \mn@doi [\apj] {10.1086/143864}, \href
  {https://ui.adsabs.harvard.edu/abs/1937ApJ....86..217Z} {86, 217}

\makeatother
\end{thebibliography}

\appendix

\section{Effect of Treating the Projected Separation as Exact} \label{appendix:rproj_test}

In the main analysis, our geometric de-projection model includes the projected separation $r_\perp$ as an observable in the likelihood with Gaussian uncertainty. The \cite{Chae2026} model instead treats $r_\perp$ as a fixed, exact quantity. To assess whether this difference contributes to the discrepancy in the inferred $\gamma$, we run an additional model variant that combines the geometric de-projection of $r_{\rm true}$ (no independent semi-major axis) with the removal of the $r_\perp$ likelihood term, so that $r_\perp$ is treated as exact. This variant is the closest analog to the \cite{Chae2026} approach within our framework.

The geometric de-projection model with $r_\perp$ uncertainty yields $\gamma = 1.56$ with a 68\% credible interval of $[1.38, 1.77]$. When $r_\perp$ is instead treated as exact, the posterior shifts to $\gamma = 1.59$ with a 68\% credible interval of $[1.40, 1.82]$. The difference is $\Delta\gamma \approx 0.03$, which is small compared to both the posterior width ($\sim$0.2) and the shift of $\Delta\gamma \approx 0.56$ caused by replacing the independent semi-major axis with the geometric de-projection. In terms of $\Gamma = \log_{10}\sqrt{\gamma}$, the two variants give $\Gamma = 0.096$ and $\Gamma = 0.101$, respectively.

\begin{figure*}
    \centering
    \includegraphics[width=0.49\columnwidth]{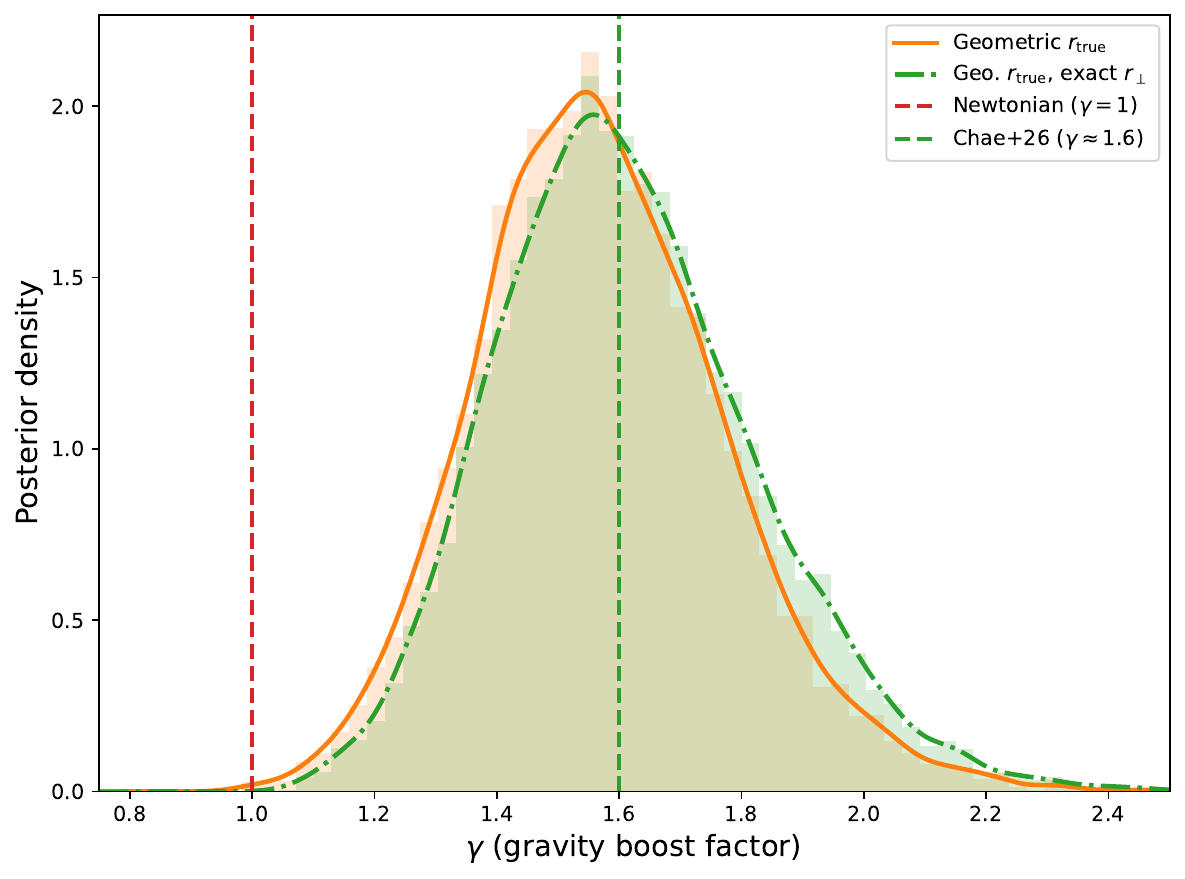}
    \includegraphics[width=0.49\columnwidth]{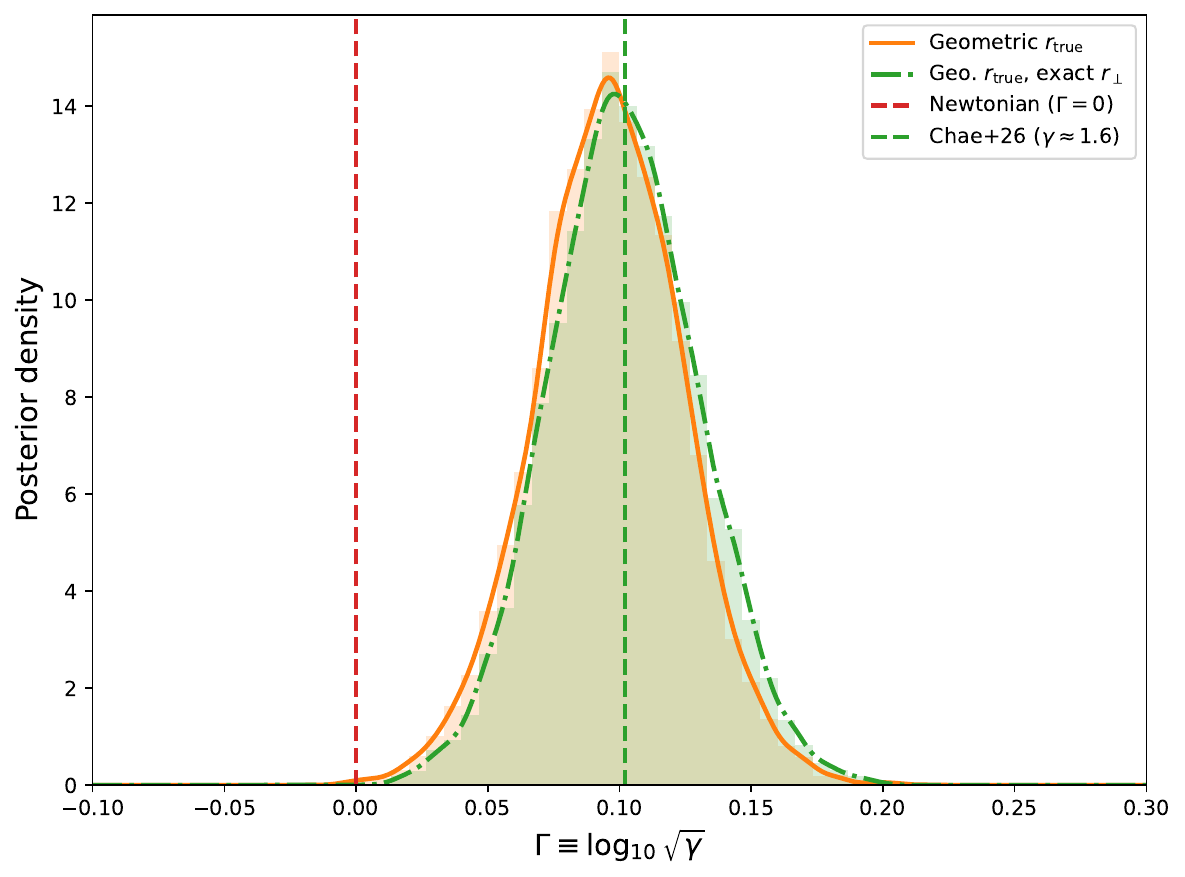}
    \caption{
    Posterior distributions comparing the effect of treating the projected separation as uncertain versus exact, both using the geometric de-projection for $r_{\rm true}$ (no independent semi-major axis).
    \emph{Left:} posterior distributions of $\gamma$. The orange solid curve shows the geometric de-projection model with $r_\perp$ entering the likelihood with Gaussian uncertainty ($\gamma = 1.56$). The green solid curve shows the same model but with $r_\perp$ treated as exact ($\gamma = 1.59$). The vertical red dashed line marks the Newtonian prediction ($\gamma = 1$) and the vertical green line shows the result from \cite{Chae2026}.
    \emph{Right:} the same posteriors expressed in terms of $\Gamma \equiv \log_{10}\sqrt{\gamma}$. The horizontal dotted line shows the flat prior $\Gamma \sim \mathcal{U}(-1,1)$.
    The two curves are nearly identical, demonstrating that the $r_\perp$ treatment has a negligible effect ($\Delta\gamma \approx 0.03$) compared to the semi-major axis parameterization ($\Delta\gamma \approx 0.56$).
    }
    \label{fig:rproj_test}
\end{figure*}

Figure~\ref{fig:rproj_test} shows the posterior distributions for both variants in $\gamma$ (left) and $\Gamma$ (right). The two posteriors overlap very well, confirming that the treatment of projected separation uncertainty is subdominant. The dominant source of the discrepancy between our baseline result and \cite{Chae2026} is the semi-major axis parameterization, not the handling of $r_\perp$.

\section{Posterior Distribution of the True Anomaly} \label{appendix:nu}

\begin{figure}
    \centering
    \includegraphics[width=0.5\columnwidth]{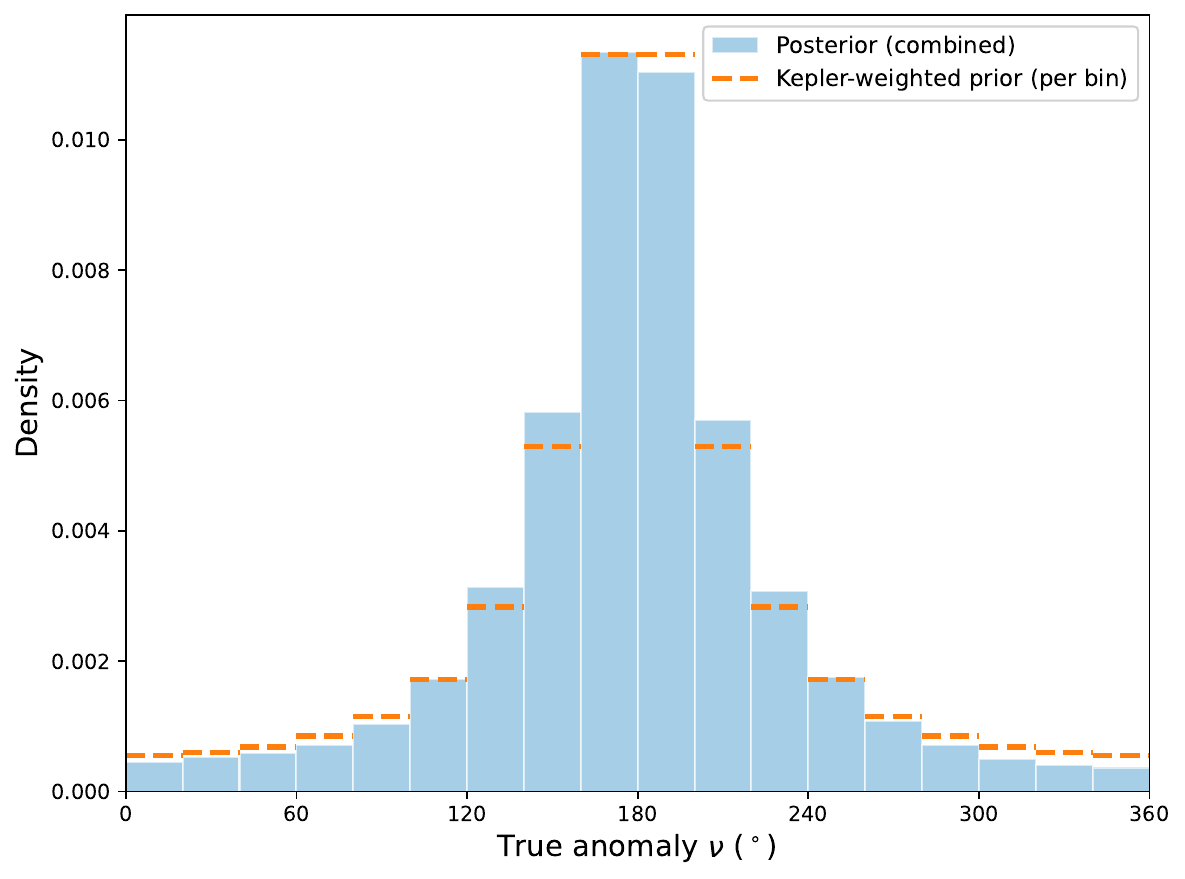}
    \caption{Population-level distribution of the posterior true anomaly $\nu$ for the 36 wide-binary systems. The blue histogram shows the composite of posterior samples pooled over all systems and MCMC draws. The orange dashed segments show the Kepler time-weighted prior for the inferred eccentricities, integrated across each bin and drawn over the full bin width so that it is directly comparable to the binned posterior. The distribution peaks near apocentre ($\nu = 180^\circ$), following the style of \citet{Chae2026} (their Figure 16).}
    \label{fig:true_anomaly}
\end{figure}

Figure~\ref{fig:true_anomaly} shows the population-level posterior distribution of the true anomaly $\nu$ for the 36 wide-binary systems, following the diagnostic of \citet{Chae2026} (their Figure 16). The composite histogram is formed by pooling all MCMC draws across all systems. The distribution peaks near apocentre ($\nu = 180^\circ$) and is consistent with the Kepler time-weighted prior computed from the inferred eccentricities (orange curve). This agreement confirms that the orbital phases are sampled in a manner consistent with the expected Keplerian behavior.

\section{Recovery of an Injected Gravity Boost} \label{appendix:injection}

Both the editor and the first referee suggested an independent check that the inference recovers the correct gravity boost from data with a known input. We therefore carry out an injection--recovery test: we build synthetic wide-binary catalogs whose relative velocities are generated for a known Newtonian value $\gamma_{\rm true}=1$, pass each catalog through the identical baseline pipeline, and compare the recovered $\gamma$ with the input. A correct procedure should return the injected value on average, with no systematic pull towards either the Newtonian value or the anomalous value reported by \citet{Chae2026}. The full procedure is summarized in Figure~\ref{fig:injection_schematic}.

\begin{figure}
    \centering
    {\begin{tikzpicture}[
    node distance=0.62cm and 1.1cm,
    box/.style={draw, rounded corners=3pt, minimum width=6.6cm, minimum height=0.62cm,
                align=center, font=\small},
    data/.style={box, fill=yellow!20, draw=black},
    drawn/.style={box, fill=blue!15, draw=black, thick},
    mstep/.style={box, fill=gray!12, draw=black},
    result/.style={box, fill=red!10, draw=black, thick},
    inp/.style={draw, rounded corners=3pt, fill=orange!18, minimum height=0.6cm,
                align=center, font=\footnotesize},
    arrow/.style={-{Stealth[length=2.5mm]}, thick, black},
]

\node[data] (real) {\textbf{Real system $j$} \;(one of 36):\\ \textit{Gaia} positions, distances, masses,\\ per-observable uncertainties $\sigma_{r_\perp},\,\sigma_{\Delta v_r},\,\sigma_{\Delta\mu}$};
\node[drawn, below=of real] (elem) {\textbf{Draw orbital elements}\\ $e\sim p(e)\!\propto\! e$,\;\; $\cos i\sim\mathcal{U}(-1,1)$,\\ $\Omega,\,\omega,\,M\sim\mathcal{U}(0,2\pi)$};
\node[mstep, below=of elem] (anchor) {\textbf{Anchor the scale:}\\ set $a$ so predicted $r_\perp$ = observed $r_\perp$};
\node[mstep, below=of anchor] (fwd) {\textbf{Forward model} (Eqs.~\ref{eq:kepler}--\ref{eq:dpm}),\\ velocity scaled by $\sqrt{\gamma_{\rm true}}$};
\node[mstep, below=of fwd] (noise) {\textbf{Project} to $r_\perp,\Delta v_r,\Delta\mu_\alpha,\Delta\mu_\delta$;\\ add Gaussian noise matched to real $\sigma$};
\node[data, below=of noise] (syn) {\textbf{Synthetic catalog} (36 systems)};
\node[mstep, below=of syn] (fit) {\textbf{Fit} with the identical baseline model\\ (broken power-law prior on $a$)};
\node[result, below=of fit] (rec) {\textbf{Recovered $\gamma$}\quad vs\quad injected $\gamma_{\rm true}=1$};

\foreach \a/\b in {real/elem, elem/anchor, anchor/fwd, fwd/noise, noise/syn, syn/fit, fit/rec}
    \draw[arrow] (\a) -- (\b);

\node[inp, right=1.0cm of fwd] (gt) {injected\\ $\gamma_{\rm true}=1$};
\draw[arrow] (gt) -- (fwd);

\node[draw, rounded corners=4pt, fill=white, align=left, anchor=north west,
      font=\footnotesize] at ($(real.north east)+(0.5,0.0)$) {
\begin{tikzpicture}[x=1cm,y=1cm]
\node[data, minimum width=0.45cm, minimum height=0.28cm] at (0,0) {};
\node[anchor=west, font=\scriptsize] at (0.45,0) {Real data};
\node[drawn, minimum width=0.45cm, minimum height=0.28cm] at (0,-0.4) {};
\node[anchor=west, font=\scriptsize] at (0.45,-0.4) {Random draw};
\node[mstep, minimum width=0.45cm, minimum height=0.28cm] at (0,-0.8) {};
\node[anchor=west, font=\scriptsize] at (0.45,-0.8) {Model step};
\node[result, minimum width=0.45cm, minimum height=0.28cm] at (0,-1.2) {};
\node[anchor=west, font=\scriptsize] at (0.45,-1.2) {Result};
\end{tikzpicture}
};

\end{tikzpicture}}
    \caption{Schematic of the injection--recovery test. Starting from each real system (yellow), we draw a full set of orbital elements (blue), set the semi-major axis so that the predicted projected separation matches the observed one, evaluate the identical forward model with the velocity scaled by $\sqrt{\gamma_{\rm true}}$ for the Newtonian input $\gamma_{\rm true}=1$, project to the four observables and add measurement noise matched to the real per-system uncertainties (grey), and analyse the resulting synthetic catalog with the identical baseline model to obtain a recovered $\gamma$ (red). The full pipeline is repeated for five independent realizations.}
    \label{fig:injection_schematic}
\end{figure}

\subsection{Generating the synthetic catalogs}

\begin{figure*}
    \centering
    \includegraphics[width=\textwidth]{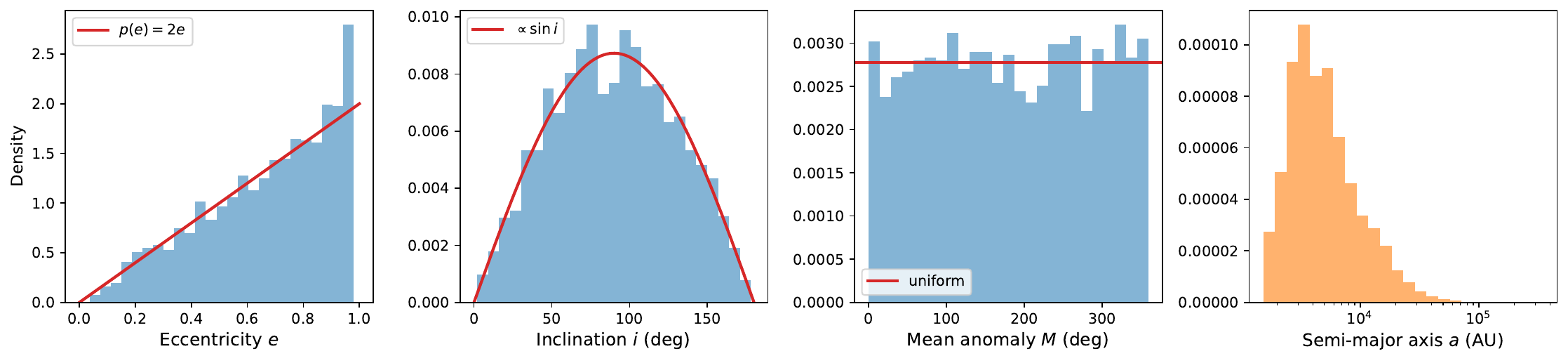}
    \caption{The generative model used to build the synthetic catalogs, pooled over 60 catalogs. The orbital elements of each synthetic binary are drawn from a thermal eccentricity distribution $p(e)\propto e$ (left, red line), an isotropic inclination distribution ($\propto\sin i$, second panel), and a uniform mean anomaly (third panel). The semi-major axis of each system (right) is then set so that its predicted projected separation matches the observed separation of the corresponding real system, which fixes the range of orbital scales shown.}
    \label{fig:injection_inputs}
\end{figure*}

Each synthetic catalog is built to reproduce the real sample system by system. For every one of the 36 binaries we retain its \textit{Gaia} sky positions, the two component distances, the stellar masses, and the measurement uncertainties on the four observables (the projected separation $r_\perp$, the radial-velocity difference $\Delta v_r$, and the two differential proper-motion components $\Delta\mu_\alpha$ and $\Delta\mu_\delta$). We then assign each synthetic binary a complete set of orbital elements: the eccentricity is drawn from a thermal distribution $p(e)\propto e$ (truncated at $e=0.98$), the inclination is isotropic ($\cos i \sim \mathcal{U}(-1,1)$), and the longitude of the ascending node, argument of periastron, and mean anomaly are uniform on $[0,2\pi)$. The semi-major axis of each synthetic binary is then fixed so that its predicted projected separation equals the observed separation of the corresponding real system. Because the projected separation scales linearly with the orbital scale at fixed angles and phase, this anchoring makes the synthetic catalog span the same range of separations as the real sample and keeps every synthetic $r_\perp$ positive. Figure~\ref{fig:injection_inputs} shows the resulting distributions of the drawn orbital elements and the anchored semi-major axes.

\begin{figure*}
    \centering
    \includegraphics[width=\textwidth]{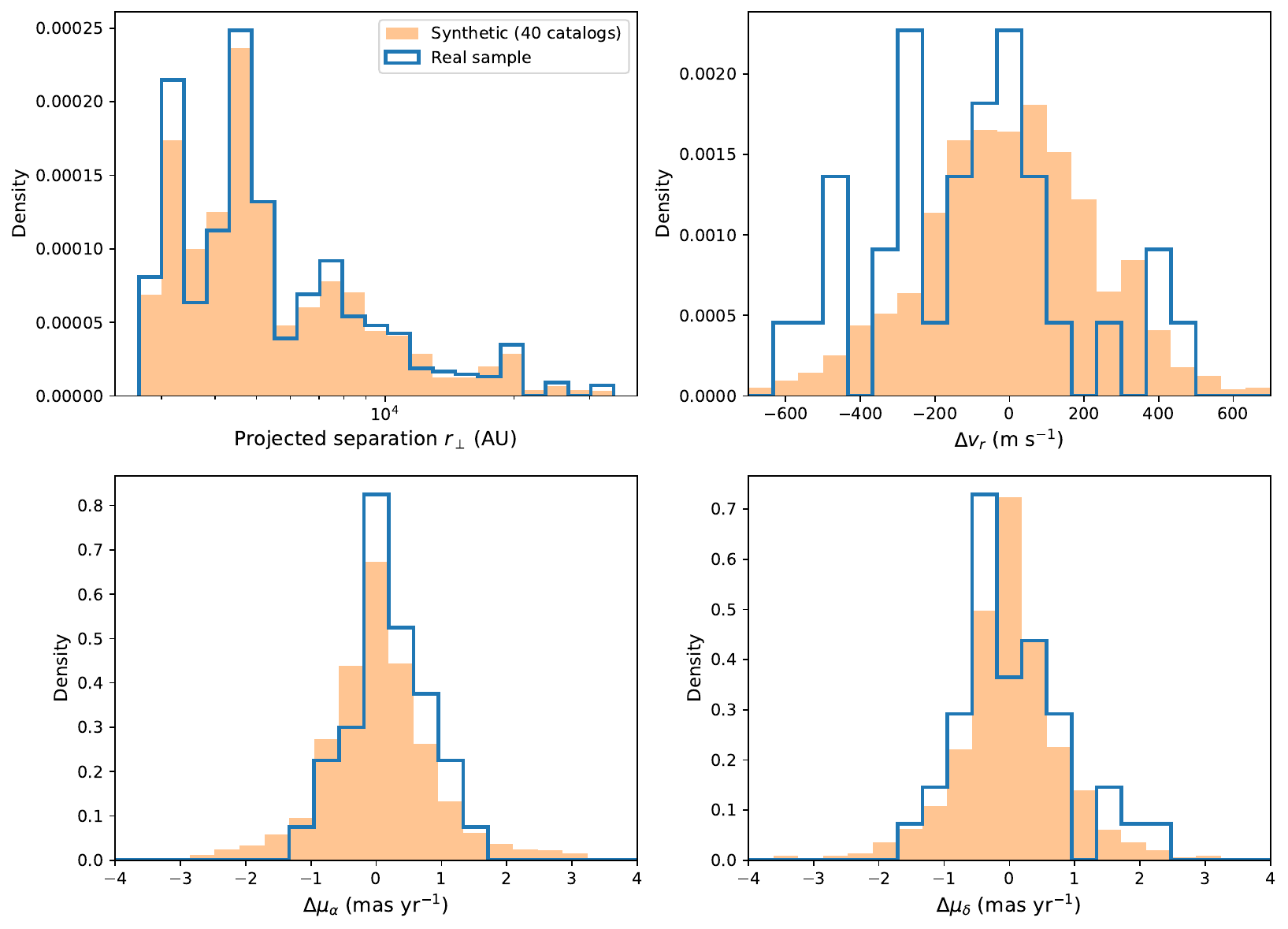}
    \caption{The synthetic catalogs reproduce the real sample. Each panel compares the distribution of one observable in the real 36-system sample (blue step histogram) with the pooled distribution from 40 synthetic Newtonian catalogs (orange filled): the projected separation $r_\perp$, the radial-velocity difference $\Delta v_r$, and the two differential proper-motion components $\Delta\mu_\alpha$ and $\Delta\mu_\delta$. The semi-major axes of the synthetic binaries are set so that their projected separations match the real systems, and the noise is drawn from the real per-system uncertainties, so the synthetic and real distributions overlap across all four observables.}
    \label{fig:injection_synthetic}
\end{figure*}

With the orbit fixed, the relative velocity of each pair is computed from the identical forward model of Section~\ref{sec:hiermodel} (Equations~\ref{eq:kepler}--\ref{eq:dpm}), with the orbital velocity scaled by $\sqrt{\gamma_{\rm true}}$, and projected onto the same four observables. Gaussian measurement noise is then added to each observable, with per-system standard deviations set to those of the corresponding real system. Each noisy synthetic catalog is analysed with exactly the same baseline model used on the real data, including the independent semi-major axis with its broken power-law prior (Equation~\ref{eq:sma_prior}) and the identical sampler settings. We generate five independent realizations at $\gamma_{\rm true}=1$, each with its own orbital-element and noise draws. The eccentricity distribution used to \emph{generate} the data is deliberately not identical to the prior used to \emph{fit} it, so the test also checks that the recovery is insensitive to the exact orbital-element prior. Figure~\ref{fig:injection_synthetic} shows that the synthetic catalogs reproduce the distributions of all four observables in the real sample.

What we should, and should not, expect from this test follows from the size of the sample. Each synthetic catalog is a single realization of 36 noisy systems, so the recovered $\gamma$ is a random quantity: its expectation over many noise realizations equals the injected value, but any one realization scatters around that value by roughly the posterior width. A single clean Newtonian catalog therefore yields a posterior that is narrower than the real-data posterior and centred near, but not exactly at, $\gamma=1$; returning exactly $1.000$ would require either noise-free data or an infinite sample, and would in fact be a coincidence. The offset of any single realization is set by its particular noise draw, not by a bias. The signature of an unbiased method is that, across realizations, the recovered values straddle the injected value with no preferred direction, and that none is pulled towards the anomalous $\gamma\approx1.6$. A biased procedure, one that produced a spurious Newtonian result or pulled a genuine anomaly towards unity, would instead return values systematically offset from the input.

\subsection{Results}

\begin{figure*}
    \centering
    \includegraphics[width=\textwidth]{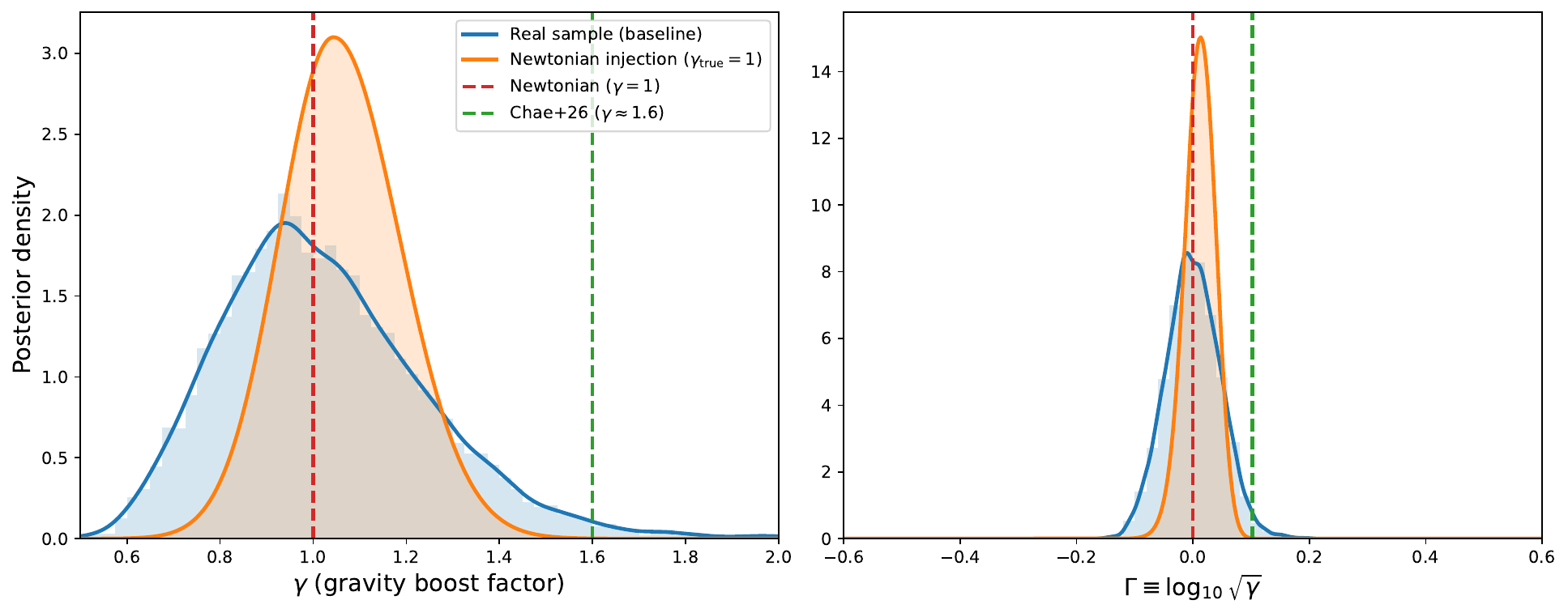}
    \caption{Injection--recovery test for a Newtonian input ($\gamma_{\rm true}=1$), shown for a single representative realization. \emph{Left:} posterior of $\gamma$; \emph{right:} the same expressed as $\Gamma\equiv\log_{10}\sqrt{\gamma}$. The orange curve is the posterior recovered from one synthetic Newtonian catalog (recovered $\gamma=1.04^{+0.14}_{-0.12}$); the blue curve is the posterior obtained from the real sample ($\gamma=1.00$). The Newtonian recovery is narrower than the real-data posterior, centred on the Newtonian prediction (red dashed line), and well separated from $\gamma\approx1.6$ reported by \citet{Chae2026} (green dashed line). The recovered values for all five realizations are listed in the text above.}
    \label{fig:injection}
\end{figure*}

Across the five Newtonian realizations we recover $\gamma = \{0.69,\,0.88,\,0.95,\,1.04,\,1.10\}$, with a median of $0.95$, a mean of $0.93$, and a realization-to-realization scatter of $0.14$; all chains are well converged ($\hat{R}\leq1.01$, bulk effective sample size above $1000$). The recovered values straddle the injected $\gamma_{\rm true}=1$ with no systematic offset, the mean lying within $0.07/\sqrt{5}$ of unity, and none approaches $\gamma\approx1.6$. Figure~\ref{fig:injection} shows a single representative realization (recovered $\gamma=1.04^{+0.14}_{-0.12}$) overlaid on the real-sample posterior: the Newtonian recovery is a clean, unimodal distribution that is narrower than the real-data posterior, sits on the Newtonian value, and is well separated from the value reported by \citet{Chae2026}. Across the five realizations the recovered values scatter about $\gamma_{\rm true}=1$ with no preferred direction; the lowest draw ($\gamma=0.69$) is the expected tail of a five-sample set rather than a systematic trend. The test therefore shows that the baseline procedure recovers a Newtonian input without bias, and that the Newtonian result obtained on the real data reflects the data rather than a limitation of the method.

\section{High-Velocity Systems and the Population-Level Result} \label{appendix:highv}

A small number of pairs in this sample have an observed relative velocity that exceeds the largest Newtonian escape velocity their measured masses and sky-projected separation allow. Because the three-dimensional separation is always at least the projected one, the escape velocity evaluated at $r_\perp$ is the most generous form of that bound, so a pair with $v_{\rm obs} > \sqrt{2 G_{\rm N} M_{\rm tot}/r_\perp}$ cannot be matched by any bound two-body Newtonian orbit. Using the ratios tabulated by \citet{Chae2026}, six of the 36 systems have a central ratio above unity, three exceed the bound at more than $2\sigma$, and one exceeds it at more than $3\sigma$. We record this explicitly because it is the sharpest point of disagreement in the recent literature on this dataset.

What such a velocity establishes is that the pair in question is not a bound two-body Newtonian system. It does not by itself identify the cause. An unresolved close companion adds an orbital reflex to the apparent relative velocity, and a radial-velocity zero-point offset between two stars observed on different instruments and at different epochs is of the same order as the excess, which for the most extreme case is $124$~m\,s$^{-1}$. An error in the adopted masses or distances of about 15 per cent also closes the gap, since the bound scales as $\sqrt{M_{\rm tot}/r_\perp}$, and a pair need not be bound at all. Modified gravity is one entry on that list rather than the only one. We note that \citet{ChaeYoon2026} make the same point in the two-dimensional case, writing that a velocity cut may remove bound binaries with boosted gravity and hidden companions alike, and that a kinematic contaminant such as a hidden close companion boosts the measured velocity.

Our inference is a population-level measurement rather than a set of per-system determinations. The gravity boost $\gamma$ is a single global parameter shared by all 36 systems, and the radial-velocity likelihood is a heavy-tailed Student-$t$, so a small number of systems lying far from the population are treated as outliers and their influence on the global parameter is bounded. This is a deliberate modelling choice, and it is the reason that a handful of contamination-prone systems do not set the result. Refitting the sample with those systems removed illustrates the point. Excluding the four systems that lie most significantly above the bound, or all six with a central ratio above unity, changes the inferred $\gamma$ by less than the width of the posterior once the truncation induced by such a velocity cut is accounted for, and in neither case does the result approach $\gamma \approx 1.6$. Fitted on their own, by contrast, the high-velocity systems prefer $\gamma \approx 4$, well above any MOND-predicted boost, which is the behaviour expected of contaminated or unbound pairs rather than of a universal modification of gravity. Readers should therefore understand the value of $\gamma$ reported here as a statement about the sample as a whole, and not as a claim that every individual pair is consistent with Newtonian gravity.

\end{document}